\documentclass[12pt, a4paper]{article}

\usepackage{natbib}
\usepackage{graphicx}
\usepackage{subfigure}

\usepackage{verbatim}
\usepackage{amssymb}
\usepackage{amsmath}
\usepackage{mathdots}
\usepackage{setspace}
\usepackage{stmaryrd}
\usepackage{subfigure}

\newcommand{\bb}{\mathcal{B}}

\DeclareMathOperator{\tr}{tr}

\DeclareMathOperator{\Grad}{Grad}
\DeclareMathOperator{\Div}{Div}

\newcommand{\be}{\begin{equation}}
\newcommand{\en}{\end{equation}}

\def\bga#1\ega{\begin{gather}#1\end{gather}} 
\def\bgas#1\egas{\begin{gather*}#1\end{gather*}}

\usepackage{authblk}

\begin{document}

\title{Morphology of residually-stressed tubular tissues: Beyond the elastic multiplicative decomposition}

\author[1,2]{P. Ciarletta}
\author[3]{M. Destrade}
\author[3,4]{A. L. Gower}
\author[2,5]{M. Taffetani}
\affil[1]{Sorbonne Universit$\rm \acute{e}$s, UPMC Univ Paris 06, CNRS, UMR 7190,
Institut Jean Le Rond d'Alembert, F-75005 Paris, France;}
\affil[2]{MOX - Politecnico di Milano, piazza Leonardo da Vinci 32, 20133 Milano, Italy}
\affil[3]{School of Mathematics, Statistics and Applied Mathematics, NUI Galway, University Road, Galway, Ireland}
\affil[4]{School of Mathematics, University of Manchester, Oxford road, Manchester, M13 9PL, UK}
\affil[5]{Mathematical Institute, University of Oxford, Oxford OX2 6GG, United Kingdom}

\date{}
\maketitle

\bigskip
\bigskip

\begin{abstract}


Many interesting shapes appearing in the biological world are formed by the onset of mechanical instability. In this work we consider how the build-up of residual stress can cause a solid to buckle. In all past studies a fictitious (virtual) stress-free state was required to calculate the residual stress. In contrast, we use a model which is simple and allows the prescription of any residual stress field.

We specialize the analysis to an elastic tube subject to a two-dimensional residual stress, and find that incremental wrinkles can appear on its inner or its outer face, depending on the location of the highest value of the residual hoop stress. 
We further validate the predictions of the incremental theory with finite element simulations, which allow us to go beyond this threshold and predict the shape, number and amplitude of the resulting creases.

\end{abstract}

\bigskip



\newpage


\section{Introduction}


The development of living materials interrelates biological processes at the molecular level and feedback mechanisms with the external environment. 
As a result, living matter is regulated by  mechano-sensing receptors (e.g. integrins, cadherins) at the cellular level, which determine the behavior at the macroscopic level. 
In particular, it is now well-acknowledged that a mechanical coupling drives the material properties of an adult tissue, which are somewhat optimized for their physiological functions \cite{bao03}. 
This mechanical feedback produces \emph{residual stresses} within the material, which we define as the self-equilibrated  stresses inside the body which persist in the absence of both external loads and geometrical constraints.
This set-up  happens for example in arteries, mechanically acting as soft thick tubes subjected to an internal blood pressure. 
Differential growth inside the artery produces an inhomogeneous residual stress, which tends to establish an optimal structural response to the internal pressure \cite{chuong1986residual}.

As the growth creates residual stress, the residual stress in turn can cause the tissue to become unstable and wrinkle or crease. One motivation for this work is to discover how much can we learn about the residual stresses from observing the formation (or avoidance) of wrinkles and creases.  

From a modeling standpoint, much work has been done by Anne Hoger and her coworkers \cite{hoger85,hoger86,hoger93,hoger96,johnson93} in the last decades to define a hyperelastic constitutive theory of soft materials with residual stresses. 
They proposed a multiplicative decomposition \cite{rodriguez,skalak96} to deal with volumetric growth in living materials, based on the framework introduced by Kroner-Lee for plasticity.  
Using a virtual state for the kinematic description of the grown material, these seminal articles demonstrated that if such a state is geometrically incompatible, then residual stresses can be calculated by considering the elastic strains which are necessary to restore compatibility of the spatial configuration. 
This approach has proved very popular and successful in the last couple of decades, inspiring an entire generation of researchers to work on the biomechanics of growth and remodeling. 
Just to mention a few applications, this theoretical framework has been extended to define the required thermo-mechanical restrictions on the stress-dependent evolution laws, and has powered the rise of morphoelastic theories, which deal with the analysis of the influence of mechanical effects on pattern selection in growing tissues.

Although Hoger's theoretical framework provides an elegant description of the residual stress distribution inside a material, its main drawback is that it refers to both an ungrown configuration and a grown virtual state which are not always accessible in experimental practice. 
In general this virtual state corresponds to the collection of stress-free states of each volume element of the initially stressed material. 
In only very few special cases can such a state be referred to as a configuration and be attained in physical practice, e.g. by cutting the material to remove the pre-stresses. 
Furthermore, recent experimental studies have demonstrated that the distribution of the residual stresses in living materials is very complex and mostly three-dimensional. 
In practice, it is not possible to release all residual stresses by making cuts along preferred directions, and thus it is rarely possible to properly identify the virtual state.

In this work we aim at extending the existing theoretical framework by working with a constitutive model which describes the distribution of  the residual stresses in living materials without introducing any virtual state or natural configuration.  
In Section \ref{section2}, we introduce the theory of initial stresses in elastic solids and propose a novel constitutive law for the strain energy function taking into account a functional dependence on both the elastic strains and the residual stresses. 
In Section \ref{section3}, we apply our constitutive theory to incompressible, hyperelastic tubular tissues. 
In particular, we use the theory of incremental deformations superposed on finite fields to study the stability of the residually-stressed configuration in circular tubes. 
The incremental boundary value problem is then solved for three general classes of residual stress distributions.  
In Section \ref{SEC:numeric} we propose a finite element implementation of the model, and perform post-buckling simulations to study the bifurcated morphology of the residually stressed configurations.  
Finally in Section \ref{section5}, we provide a critical discussion on the results of this work together with some concluding remarks.


\section{Hyperelastic theory of residual stresses}
\label{section2}


Let $\bb_{\boldsymbol \tau}$ be the region occupied by a soft material in its reference configuration. 
Denoting by ${\bf X}$ the material position in  $\bb_{\boldsymbol \tau}$, we assume that the body is subjected to a residual stress ${\boldsymbol \tau}= {\boldsymbol \tau}({\bf X})$ in this configuration. 
It must be symmetric to satisfy the balance of angular momentum, while the balance of linear momentum in quasi-static conditions reads:
\begin{equation}
\Div {\boldsymbol \tau} = {\bf 0} \qquad {\rm in} \quad \bb_{\boldsymbol \tau},
\label{eqres}
\end{equation}
where $\Div$ is the material divergence.
We apply the zero-traction boundary conditions
\begin{equation}
{\boldsymbol \tau} {\bf N} = {\bf 0} \qquad {\rm in} \quad
\partial \bb_{\boldsymbol \tau}, \label{bcref}
\end{equation}
where $\partial \bb_{\boldsymbol \tau}$ is the boundary of the body $\bb_{\boldsymbol \tau}$, and ${\bf N}$ is its outer unit normal. 

An important consequence of Eqs.\eqref{eqres} and~\eqref{bcref} is that the residual stress ${\boldsymbol \tau}$ must be inhomogeneous and have zero average over the volume in $\bb_{\boldsymbol \tau}$, 
\begin{equation}
\int_{\bb_{\boldsymbol \tau}} {\boldsymbol \tau} dv = {\bf 0},  \label{average}
\end{equation}
which can be shown by applying a version of the mean value theorem \cite{hoger85}.

Let us now consider that the body can be elastically deformed to a new configuration $\bb$, so that a mapping $\boldsymbol \chi: \bb_{\boldsymbol \tau} \rightarrow\bb$  defines the spatial position ${\bf x}= \boldsymbol{\chi}({\bf X})$ and ${\bf F}=\partial \boldsymbol\chi/\partial{\bf X}$ is the deformation gradient.  
Assuming that the body is perfectly elastic, it is possible to define a strain energy density $\Psi$ per unit of reference volume. 
Since the body is residually stressed, $\Psi$ must be necessarily inhomogeneous, so we assume the functional dependence
$\Psi=\Psi({\bf F}, {\boldsymbol \tau})$.
Assuming that the body is incompressible, i.e. $J= \det {\bf F}= 1$, the first Piola-Kirchoff ${\bf P}$ and Cauchy ${\boldsymbol \sigma}$ stress tensors are
\begin{equation}
{\bf P}= \frac{\partial \Psi}{\partial {\bf F}}({\bf F},
{\boldsymbol \tau})-p{\bf F}^{-1}, 
\qquad {\boldsymbol \sigma}={\bf F P}, \label{stresses}
\end{equation}
where $p$ is the Lagrange multiplier associated with the internal constraint of incompressibility. 
Evaluating Eq.\eqref{stresses} in $\bb_{\boldsymbol \tau}$ by setting $\boldsymbol F= \boldsymbol I$, we derive a connection for the residual stress,
\begin{equation}
{\boldsymbol \tau}= \frac{\partial \Psi}{\partial {\bf F}}({\bf
I}, {\boldsymbol \tau})-p_{\boldsymbol \tau} {\bf I}, \label{res-stress}
\end{equation}
where $\bf{I}$ is the identity matrix and $p_{\boldsymbol \tau}$ is the value of $p$
in $\bb_{\boldsymbol \tau}$.

The presence of residual stress generally introduces anisotropy in the material response, but if we assume that there is no other source of anisotropy then the strain energy density $\Psi$ should be objective. That is, $\Psi$ should be invariant under rigid body rotations of the current configuration, from which we can conclude that $\Psi$ depends on both the left Cauchy-Green strain tensor ${\bf C}={\bf F}^T{\bf F} $ and ${\boldsymbol \tau}$. Note that as ${\boldsymbol \tau}$ is based in the reference $\bb_{\boldsymbol \tau}$ it does not change by rotating the current configuration. Further by taking rigid body rotations of the reference body, it can be shown~\cite{shams} that $\Psi$ depends only on the principal invariants $I_j$ and $I_{\sigma j}$ $(j=1,2,3)$ of  ${\bf B}={\bf F}{\bf F}^T$ and ${\boldsymbol \tau}$, respectively,  
\begin{align}
& I_1= \tr {\bf B}, && I_2=\tfrac{1}{2}[(I_1^2 -\tr({\bf B}^2)], && I_3= \det {\bf B} \label{I13}, \\
& I_{\boldsymbol \tau 1}= \tr {\boldsymbol \tau}, && I_{\boldsymbol \tau 2}= \tfrac{1}{2}[(I_{\boldsymbol \tau 1}^2 -\tr({\boldsymbol \tau}^2]), &&  I_{\boldsymbol \tau 3}= \det {\boldsymbol \tau}, \label{Isigma13}
\end{align}
and on the combined invariants $J_i$ $(i=1,..,4)$,
\begin{equation}
J_1= \tr ({\boldsymbol \tau}{\bf C}), \qquad 
J_2= \tr ({\boldsymbol \tau} {\bf C}^2), \qquad 
J_3= \tr ({\boldsymbol \tau}^2 {\bf C}), \qquad 
J_4= \tr ({\boldsymbol \tau}^2 {\bf C}^2).
\label{Imix}
\end{equation}

Writing $\Psi=\Psi(I_j, I_{\boldsymbol \tau j}, J_i)$, Eq.\eqref{stresses} for the Cauchy stress of an incompressible material becomes
\begin{multline}
{\boldsymbol \sigma}
=2\frac{\partial \Psi}{\partial I_1}{\bf B}
+2\frac{\partial \Psi}{\partial I_2}(I_1{\bf B}-{\bf B}^2)-p {\bf I}\\ 
+2\frac{\partial \Psi}{\partial J_1}{\boldsymbol \Sigma} 
+2\frac{\partial \Psi}{\partial J_2}({\boldsymbol \Sigma}{\bf B}
+{\bf B}{\boldsymbol \Sigma}) +2\frac{\partial \Psi}{\partial J_3}{\boldsymbol \Sigma}{\bf B}^{-1}{\boldsymbol \Sigma}+2\frac{\partial \Psi}{\partial J_4}({\boldsymbol \Sigma}{\bf B}^{-1}{\boldsymbol \Sigma}{\bf B}+{\bf B}{\boldsymbol \Sigma}{\bf B}^{-1}{\boldsymbol \Sigma}),
\label{sigmagen}
\end{multline}
where ${\boldsymbol \Sigma} \equiv {\bf F} {\boldsymbol \tau} {\bf F}^T$.  
Recalling the required connection for the residual stresses in Eq.\eqref{res-stress}, in the reference configuration, i.e. for
${\bf F}={\bf I}$, the following conditions must hold, 
\begin{equation}
2\frac{\partial \Psi}{\partial I_1}+4\frac{\partial \Psi}{\partial I_2}-p_{\boldsymbol \tau}=0, \qquad 
2\frac{\partial \Psi}{\partial J_1}+4\frac{\partial \Psi}{\partial J_2}=1, \qquad 
\frac{\partial \Psi}{\partial J_3}+2\frac{\partial \Psi}{\partial J_4}=0. \label{nc}
\end{equation}
 
Another constraint for the choice of the strain energy for residually stressed materials is that $\Psi$ should have the same functional form for any configuration (as long as the deformation gradient $\bf F$ considered is elastic). 
This requirement results in a restriction called the \emph{initial stress symmetry}, see Gower \emph{et al.} \cite{gower15} for further details on this constitutive restriction.

In order to study the influence of residual stress on wave propagation, azimuthal shear, and torsion, Shams \emph{et al.} \cite{shams} and Merodio \emph{et al.} \cite{merodio13} proposed the following prototype constitutive equations
\begin{align}
&\Psi = \tfrac{1}{2}\mu(I_1-3) + \tfrac{1}{2}(I_6-I_{\boldsymbol \tau 1}) +\tfrac{1}{2}\overline{\mu} (I_6-I_{\boldsymbol \tau 1})^2,\notag \\
&\Psi = \tfrac{1}{2}\mu(I_1-3) + \tfrac{1}{4}(I_5-I_{\boldsymbol \tau 1}), \notag \\
& \Psi = \tfrac{1}{2}\mu(I_1-3) + \tfrac{1}{2}(I_6-I_{\boldsymbol \tau 1}),
\end{align}
respectively, where $\mu$, $\overline \mu$ are material constants.
Although they satisfy the conditions in Eq.\eqref{nc}, these candidates ignore the contribution of the invariants $J_1, J_2, J_3, J_4$, coupling the elastic deformation to the residual stresses, which is difficult to explain physically. 
Moreover, it turned out that they do not respect the initial stress symmetry \cite{gower15}. 

An original approach is to take advantage of  the existence of a virtual stress-free state, yet circumventing the need to actually define it from a kinematic viewpoint, as is done in Ref.\cite{gower15}. 
There, the following strain energy density was found to satisfy both Eq.\eqref{nc} and the initial stress symmetry, 
\be
  \Psi= \Psi(I_1, J_1,I_{\boldsymbol \tau 1},I_{\boldsymbol \tau 2}, I_{\boldsymbol \tau 3} ) = \tfrac{1}{2} \left(J_1+ \tilde{p} I_1 -3\mu \right),
  \label{engen}
\en
where $\mu>0$ is a material parameter, and $\tilde{p}=\tilde{p}(I_{\boldsymbol \tau 1},I_{\boldsymbol \tau 2}, I_{\boldsymbol \tau 3} )$ is a complicated function for the invariants of $\boldsymbol \tau$.
It is worth noticing that in the absence of residual stress,  $\Psi$ is the classical neo-Hookean strain energy function. 
Thus, Eq.\eqref{engen} represents the general extension of the neo-Hookean strain energy function for a residually stressed material, resulting in a function of only five of the nine independent invariants of  ${\bf C}$ and ${\boldsymbol \tau}$.
It is therefore expected that a functional dependence on the combined invariants $J_2, J_3, J_4$ would represent natural hyperelastic models of higher order.


\section{Soft tubes under plane residual stresses:  stability analysis}
\label{section3}



\subsection{Plane residual stress} 


From now on we consider a soft hollow cylinder in the residually-stressed reference configuration $\bb_{\boldsymbol \tau}$, indicating by $R_i$ and $R_o$ its inner and outer radii, respectively. 
Using a cylindrical coordinate system, the kinematics of the deformation can be defined by a mapping $\boldsymbol{\chi}$ bringing the material point ${\bf X} = {\bf X}(R,\Theta,Z)$ to the spatial position ${\bf x} = {\bf x}(r,\theta,z)= \boldsymbol\chi({\bf X})$ in the deformed configuration, where $(R, \Theta, Z)$ and $(r,\theta,z$) are the coordinates along the
orthonormal vector bases $({\bf E}_R, {\bf E}_\Theta, {\bf E}_Z)$ and $({\bf e}_r, {\bf e}_\theta, {\bf e}_z)$, respectively.

We consider that the cylinder behaves as a residually stressed neo-Hookean material Eq.\eqref{engen}, and assume a plane strain condition.
This assumption simplifies the expression for $\tilde{p}$ greatly \cite{gower15}, as now it is given by the root of the quadratic 
\be
 \tilde p^2 + \tilde p I_{\tau 1}  + I_{\tau 3}  -\mu^2  =0.
\en

Accordingly, the strain energy function of the pre-stressed body in plane strain conditions reads
 \be
 \Psi = \Psi(I_1, J_1,I_{\tau 1}, I_{\tau 3} ) =
\tfrac{1}{2} J_1+ \tfrac{1}{4}\left(\sqrt{ I_{\tau 1}^2 + 4 (\mu^2 - I_{\tau 3})}-I_{\tau 1}\right) I_1 -\mu,
\label{en2d} 
\en
since we must discard the negative root of $\tilde{p}$ to ensure the positiveness of the strain energy function. 
From Eqs.(\ref{sigmagen}) and (\ref{en2d}), the Cauchy stress for the residually stressed tube reads
\be
{\boldsymbol \tau}= \tfrac{1}{2}\left(\sqrt{ I_{\tau 1}^2 + 4 (\mu^2 - I_{\tau 3})}-I_{\tau 1}\right){\bf B} - \tilde p {\bf I}+{\boldsymbol \Sigma},
\label{sigma2d}
\en
with $p_{\boldsymbol \tau}= \tilde p({\bf F}= {\bf I})= \frac{1}{2}\left(\sqrt{ I_{\tau 1}^2 + 4 (\mu^2 - I_{\tau 3})}-I_{\tau 1}\right)$ from Eq.\eqref{nc}${}_1$.
Note that here and hereafter, tensors are two-dimensional, restricted to ${\bf E}_\alpha \otimes {\bf E}_\beta$ where $(\alpha, \beta) = (R,\Theta)$ in $\bb_{\boldsymbol \tau}$, and to ${\bf e}_i \otimes {\bf e}_j$ where $(i, j) = (r,\theta)$ in the current configuration $\bb$.


\subsection{Residual stress fields for the hollow cylinder}


For the residually-stressed hollow cylinder in its reference configuration $\mathcal B_r$, the equilibrium equations for the residual stress are
\begin{align}
 \left.
\begin{array}{c l}
     & \displaystyle  \frac{\partial}{\partial R} \left( R^2  {{\tau}}_{\Theta R}  \right)  + R \frac{\partial {{\tau}}_{\Theta \Theta}}{\partial \Theta} = 0, \\[6pt]
    & \displaystyle \frac{\partial}{\partial R} \left(R {{\tau}}_{RR} \right) + \frac{\partial {{\tau}}_{R\Theta}}{\partial \Theta} - {{\tau}}_{\Theta \Theta}= 0,
\end{array}\right \}
\; \textrm{ for }  R\in[R_i,R_o], \label{eq-plane}
\end{align}
complemented by the traction-free boundary conditions at the inner and outer radii,
\begin{align}
    & {{\tau}}_{RR} =  {{\tau}}_{R\Theta} = 0 \; \quad \textrm{ for  } R = R_i, \quad R = R_o,
    \label{bc-plane}
\end{align}

We can write the general solution for Eq.\eqref{eq-plane} with the Airy stress function $\varphi=\varphi(R,\Theta)$, i.e. defining the residual stress components as
 \be {{\tau}}_{RR} = \frac{1}{R}\varphi_{,R} + \frac{1}{R^2} \varphi_{,\Theta\Theta}, \quad 
 {{\tau}}_{,R\Theta} = \frac{1}{R^2} \varphi_{,\Theta} - \frac{1}{R} \varphi_{,\Theta R}, \quad
 {{\tau}}_{\Theta \Theta} = \varphi_{,RR}, 
 \label{airy} 
 \en
where the comma denotes partial differentiation. 
Eq.\eqref{airy} allows an easy definition of different classes of self-equilibrated residual stresses for the tube, by simply imposing a functional dependence $\varphi(R,\Theta)$ respecting the boundary conditions in Eq.\eqref{bc-plane} and periodicity in $\Theta$ over $2\pi$. 
This is a simpler approach than the one proposed by Rodriguez \emph{et al.} \cite{rodriguez}, which needed the \textit{a priori} definition of the virtual state of the material for describing the mapping to the pre-stressed material configuration,
thus requiring the derivation of the residual stress components and the \textit{a posteriori}  check of the required equilibrium conditions.

When we take $\varphi = \varphi(R)$ only, we obtain a residual stress with diagonal terms only and the physical fields are axi-symmetric, which simplifies the analysis. 
Then 
\be
{{\tau}}_{RR} = \frac{1}{R}f(R), \qquad 
{{\tau}}_{\Theta \Theta} = f'(R), \qquad {\rm with}\quad f(R_i)=f(R_o)=0, \label{airy2} 
\en
where $f(R) \equiv \varphi'(R)$ can be regarded as a \emph{stress potential}.

In this paper we consider three different residually stressed states in turn, as defined by the following variations for $f(R)$,
\begin{align}
 \left\{
\begin{array}{c l}
     & \displaystyle  (a) \quad f(R)=\alpha  \mu  R (R-R_i)  (R-R_o) /R_i^2,\\
      & \displaystyle (b) \quad f(R)=\alpha  \mu R  \ln(R/R_i)\ln(R/R_o), \\
      & \displaystyle (c) \quad f(R)=\alpha  \mu R  \left(\text{e}^{R/R_i}-1\right)\left(\text e^{R/R_o} - 1\right),
      \end{array}\right.
\label{fconst}
\end{align}
where $\alpha$ is a non-dimensional measure of the residual stress amplitude.
 
\begin{figure}[h!]\centering
\subfigure[(a)][parabolic]{\includegraphics[width=0.32\textwidth]{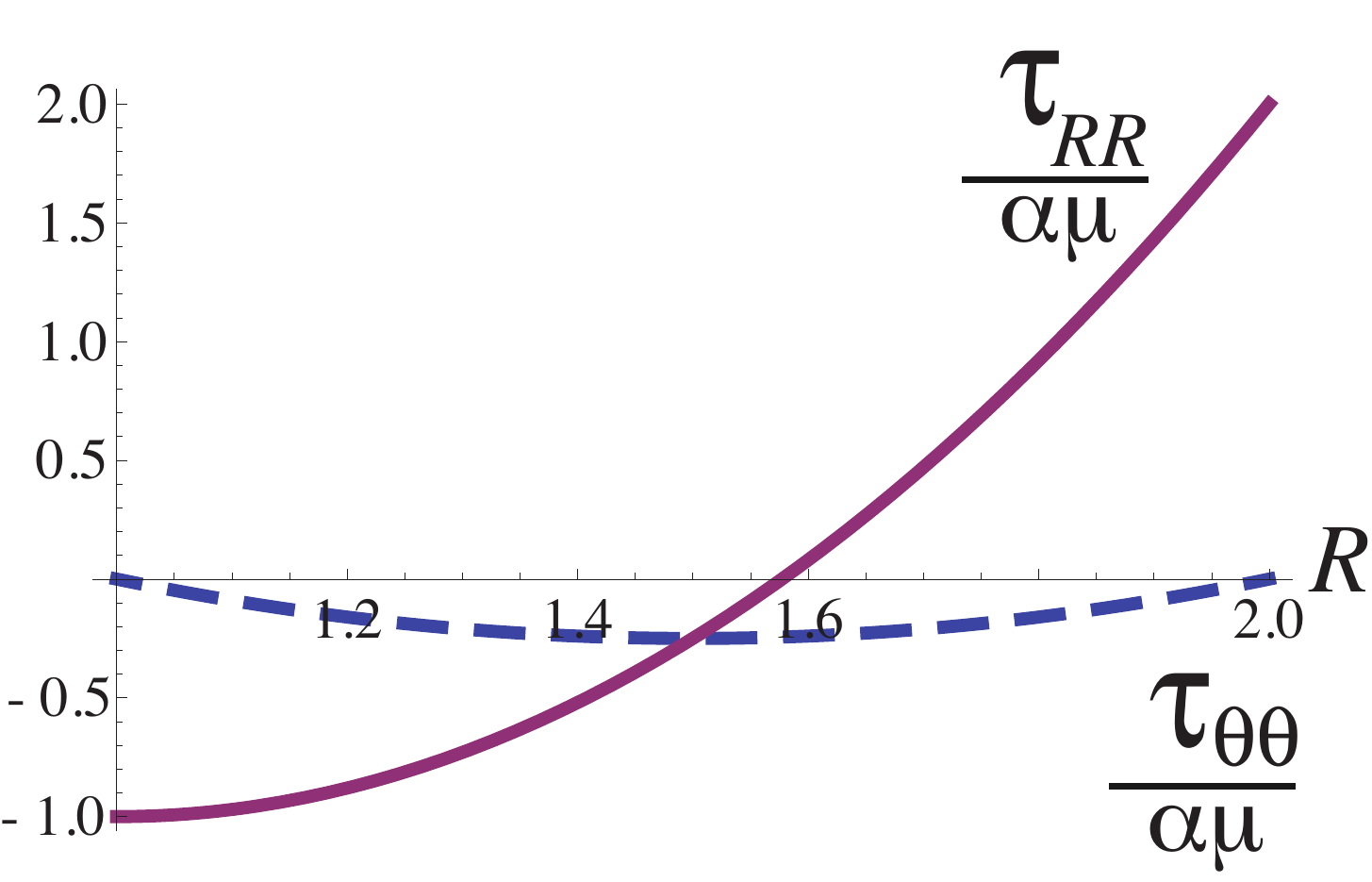}}
\subfigure[(b)][logarithmic]{\includegraphics[width=0.32\textwidth]{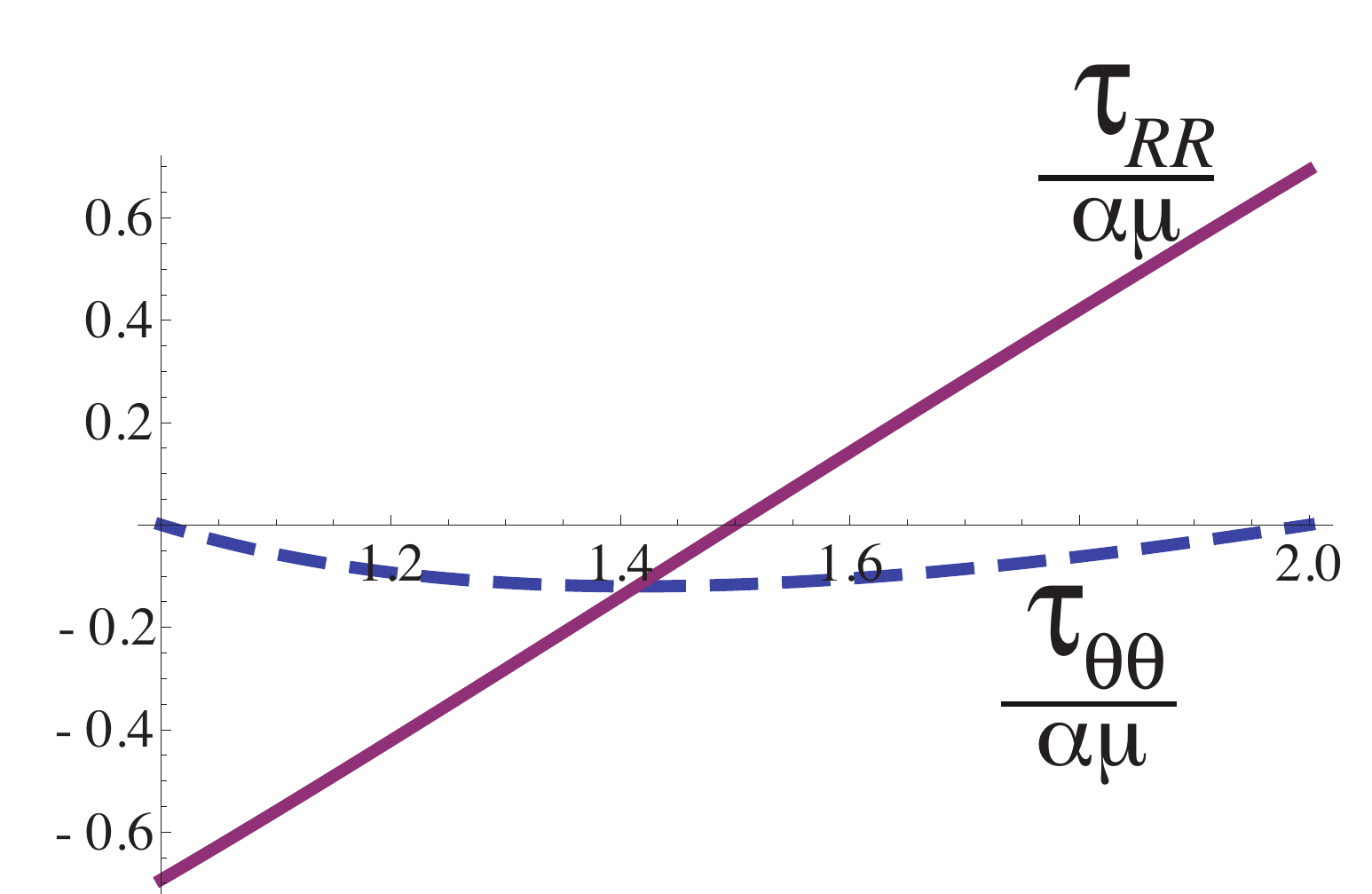}}
\subfigure[(c)][exponential]{\includegraphics[width=0.32\textwidth]{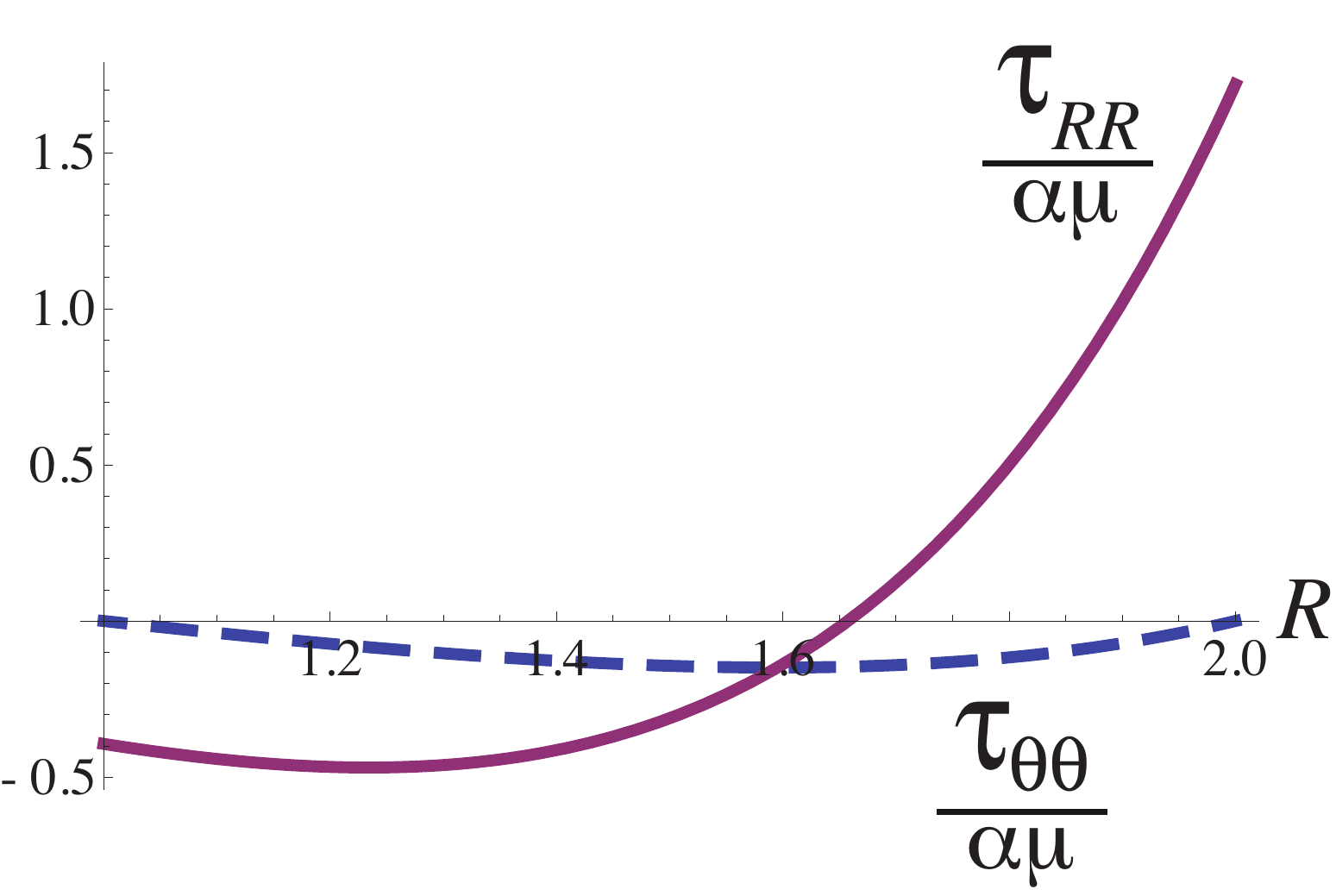}}
\caption{Radial (solid purple lines) and hoop (dashed blue lines) residual stresses for a hollow cylinder
with $R_i=1$, $R_o=2$ and the stress potential in Eq.(\ref{fconst}a)
(left), Eq.(\ref{fconst}b) (center) and Eq.(\ref{fconst}c)
(right).}\label{figres-str}
\end{figure}

Figure \ref{figres-str} depicts the corresponding radial and hoop residual stress variations through the thickness when $R_i=1$, $R_o=2$, normalized with respect to $\alpha \mu$.
When $\alpha>0$, the radial stress increases from the inner to the outer face, and it decreases when $\alpha<0$. 
We call these situations \emph{tensile} and \emph{compressive residual stress}, respectively.
As the magnitude $|\alpha|$ increases, the stress difference between the inner and the outer face can be so large as to de-stabilise the tube, as we see in the next section.



\subsection{Incremental equations}


We first investigate the stability of the residually stressed tube with the method of incremental deformations superposed on a finite field \cite{ogden1997non}, here a finite residual stress in contrast to other studies with a finite pre-strain. 
We perturb the residually-stressed, axi-symmetric reference configuration by applying a two-dimensional incremental displacement vector
${\bf u}$, expressed as 
\begin{equation}\label{incrvect}
{\bf u} = u(R,\Theta)\textbf{E}_R+v(R,\Theta)\textbf{E}_{\Theta},
\end{equation}
where $u$, $v$ are the incremental radial and hoop displacement fields, respectively.
The spatial displacement gradient associated with the incremental deformation, $\boldsymbol{\Gamma}=\Grad{\bf u}$, reads
\begin{equation}
\boldsymbol{\Gamma}=
\begin{bmatrix}
u_{,R} & (u_{,\Theta}-v)/R  \\
v_{,R} & (v_{,\Theta}+u)/R
\end{bmatrix},
\label{eqn:Gamma}
\end{equation}
while the incremental constraint of incompressibility is
\begin{equation}
\rm tr \ {\boldsymbol \Gamma}=0. 
\label{inc}
\end{equation}

The components of ${\bf s}$, the linearised nominal stress on the reference configuration $\bb_{\boldsymbol \tau}$, are
\begin{equation}\label{essepunto1}
s_{ij}= \mathcal{A}_{0ijkl}\Gamma_{lk}+p_{\boldsymbol \tau} \Gamma_{ij}-q_{\boldsymbol \tau} \delta_{ij}, 
\end{equation}
for $i, j= R, \Theta$, where $q_{\boldsymbol \tau}$ is the increment of the Lagrange multiplier $p_{\boldsymbol \tau}$ and $\boldsymbol{\mathcal{A}_0}$ is the fourth-order tensor of instantaneous elastic moduli.
Following Shams \emph{et al.} \cite{shams}, we express its components for a residually stressed material as
\begin{equation}\label{A10} 
{\mathcal{A}_{0iklj}} 
= \dfrac{\partial^2 \Psi}{\partial F_{k\alpha}\partial F_{j\beta}}
= 2 \delta_{jk} \delta_{il}\Psi_{1} + \delta_{jk}{\tau}_{il},
\end{equation}
where we take $\mathbf F = \mathbf I$ after differentiation, $\delta$ is the Kronecker delta and 
\begin{equation}
\Psi_{1} \equiv \frac{\partial \Psi}{\partial I_1} = \frac{1}{4}\sqrt{4\mu^2 + \left[\frac{1}{R} f(R) - f'(R)\right]^2} - \frac{1}{R}f(R) - f'(R).
\end{equation}

The incremental equations of equilibrium are
\begin{equation}
{\rm Div} \, \mathbf{s} = \mathbf 0, 
\label{incsigma}
\end{equation}
whilst the vanishing of the incremental traction at the free surface reads
\begin{equation}
\textbf{s}^T \textbf{E}_R= \mathbf{0} \qquad
 {\rm at} \quad R= R_i, R_o.
 \label{incbc}
\end{equation}


\subsection{Stroh formulation and surface impedance method}


Assuming a cosine variation on the faces of the tube $u(R,\Theta) = U(R) \cos(m \Theta)$, and then using Eqs.\eqref{essepunto1} and \eqref{eqn:Gamma} we reach following expressions for the incremental displacement and stress field,
\begin{equation}\label{s4}
\begin{array}{ll}
[ u(R,\Theta), s_{RR}(R,\Theta), q(R, \Theta) ] = [ U(R),  S_{RR}(R), Q(R) ] \cos(m\Theta ),\\
  {[ v(R,\Theta)  , s_{R \Theta} (R,\Theta)]} =  [ V(R),   S_{R \Theta}(R)] \sin(m\Theta),
\end{array}
\end{equation}
where the integer number $m$ is the circumferential wavenumber, and the amplitudes $U,V,S_{RR},S_{R \Theta}$ are four scalar functions of $R$ only. 
We now rewrite the governing equations in a Stroh \cite{stroh} formulation: a system with many favourable properties to solve boundary value problems. 
The first line of the Stroh form Eq.\eqref{stroh} below is just Eq.\eqref{inc} reordered.  The second line is a rewrite of Eq.\eqref{essepunto1} for the component $S_{R \Theta}$. Then we use the first two lines of Eq.\eqref{stroh} to substitute $U'$ and $V'$ in terms of $U$, $V$, $S_{RR}$ and $S_{R \Theta}$ into Eq.\eqref{incsigma}, from which we get the third and fourth line of Eq.\eqref{stroh}, resulting in 
\begin{equation}\label{stroh}
\dfrac{d}{dR} \begin{bmatrix}
\boldsymbol{\mathsf{U}}\\
R\boldsymbol{\mathsf{S}}
\end{bmatrix}  = 
\dfrac{1}{R}
\begin{bmatrix}
 \textbf{G}_1&\textbf{G}_2\\
\textbf{G}_3 & -\textbf{G}_1^T
\end{bmatrix}
\begin{bmatrix}
\boldsymbol{\mathsf{U}}\\
R\boldsymbol{\mathsf{S}}
\end{bmatrix} 
\quad \text{with}\quad \left\{
\begin{array}{ll}
\boldsymbol{\mathsf{U}}(R)=[U(R),V(R)]^T,\\
\boldsymbol{\mathsf{S}}(R)=[S_{RR}(R),S_{R\Theta}(R)]^T,
\end{array}
\right.
\end{equation} 
where the sub-blocks of the  {Stroh matrix} have the following components 
\begin{align}
& \textbf{G}_1= 
\begin{bmatrix}
-1 & -m{}^{check\text{\footnotemark} } \\
\frac{2mR\Psi_{1}}{f+2R\Psi_{1}} & \frac{2R\Psi_{1}}{f+2R\Psi_{1}}  
\end{bmatrix}, 
\qquad
\textbf{G}_2= 
\begin{bmatrix}
 0 & 0 \\
 0 & \frac{R}{f+2R\Psi_{1}} 
 \end{bmatrix},
\notag \\[12pt]
& \textbf{G}_3
= \begin{bmatrix} 
8  \Psi_{1}  + (1 + m^2)  f' + \frac{f [f + 2 (1 + m^2) R \Psi_{1}]}{R(f+ 2 R \Psi_{1})} & m \left[8 \Psi_{1} + 2f' + \frac{f (f + 4 R \Psi_{1})}{R (f + 2 R \Psi_{1})}\right]\\
 m \left[8 \Psi_{1} + 2f' + \frac{f (f + 4 R \Psi_{1})}{R (f + 2 R \Psi_{1})} \right] &   8 m^2  \Psi_{1} + (1 + m^2) \left(f'+\frac{f}{R}\right) - \frac{f^2}{R( f + 2 R \Psi_{1})}
\end{bmatrix},
\end{align}
\footnotetext{Please double check if this should be $-m$, this was previously $m$ in Pasquale's version. If it was wrong, did it effect the numerical results?}
which can be found be specialising the general expressions of Refs.\cite{DeNC09, DeOM10} to the present context.
Here we substituted $p_{\boldsymbol \tau} = \mathcal A_{01212}- \tau_{RR}$ by using Eq.~\eqref{nc}${}_1$ and Eq.~\eqref{A10}, and substituted $Q$ by using Eq.\eqref{essepunto1} with $i=j=R$.

We can solve numerically the boundary value problem formed by Eqs.\eqref{incbc} and \eqref{stroh} in a robust manner by adopting the \emph{impedance matrix
method}. 
Following Destrade \emph{et al.} \cite{DeNC09} we introduce a functional relation between the incremental traction and the displacements vectors as
\begin{equation}
 R \, \boldsymbol{\mathsf{S}}(R)= {\bf Z}(R) \,  \boldsymbol{\mathsf{U}}(R), \label{impZ}
\end{equation}
where ${\bf Z}$ is a \textit{surface impedance matrix}. 
Substituting Eq. \eqref{impZ} into Eq.\eqref{stroh}, we derive the following differential Riccati equation for $\mathbf Z$,
\begin{equation}\label{eqZ}
\dfrac{d}{dR}\textbf{Z}=\dfrac{1}{R}\left(\textbf{G}_3-\textbf{G}_1^T\textbf{Z}-\textbf{Z}\textbf{G}_1-\textbf{Z}\textbf{G}_2\textbf{Z}\right),
\end{equation}
It must be integrated numerically from the initial condition ${\bf Z}={\bf Z}(R_i)={\bf 0}$ (or, equivalently, ${\bf Z}={\bf Z}(R_o)={\bf 0}$), to the target condition,
\begin{equation}
\det {\bf Z}(R_o)= {\bf 0} \qquad (\det {\bf Z}(R_i)= {\bf 0}, {\rm respectively}).
\label{bcZ}
\end{equation}
Also in general $\textbf{Z}=\textbf{Z}^T$, see Shuvalov \cite{shuvalov03}.

Once  $f$, $\Psi$, $R_o$ and $R_i$ are prescribed for a given tube, we adjust the remaining parameter $\alpha$, proportional to the amplitude of the residual stress, until we meet the target Eq.~\eqref{bcZ}.
Once $\alpha$ is determined, we can integrate the first line of Eq.\eqref{stroh}, i.e.
\begin{equation}
\dfrac{d\mathbf U}{dR} = \dfrac{1}{R}\mathbf G_1 \mathbf U + \dfrac{1}{R}\mathbf G_2 \mathbf{ZU},
\end{equation}
simultaneously with Eq.\eqref{eqZ} to compute the incremental displacement field throughout the thickness of the tube wall.

The numerical method for solving the initial value problem given by Eqs.(\ref{eqZ}) and (\ref{bcZ}) for the three potential stress functions in Eq.\eqref{fconst} is presented in the next section.


\subsection{Numerical results on wrinkling}


The Hamiltonian structure and algebraic properties of the Stroh matrix yield a robust numerical procedure to determine when wrinkles appear either of the faces of the residually stressed tube \cite{fu, shuvalov03}. 
Here we find the unique, symmetric, semi-definite solution of the differential Riccati equation for ${\bf Z}$ in Eq.\eqref{eqZ} by numerical integration using the software {\textit{Mathematica}}  (Wolfram Inc., version 10.1, Champaign, IL)  from the initial zero value to a target condition given by the boundary condition in Eq.\eqref{bcZ}.

We consider in turn the three expressions for $f(R)$ in Eq.\eqref{fconst}. 
In each case we find the critical value $\alpha > 0$ ($\alpha <0$) for wrinkles to appear on the inner face (outer face), under tensile (compressive) residual stress. 

\begin{figure}[!htb]
\centering
{\centering
\subfigure[tensile residual stress]{\includegraphics[width=0.46\textwidth]{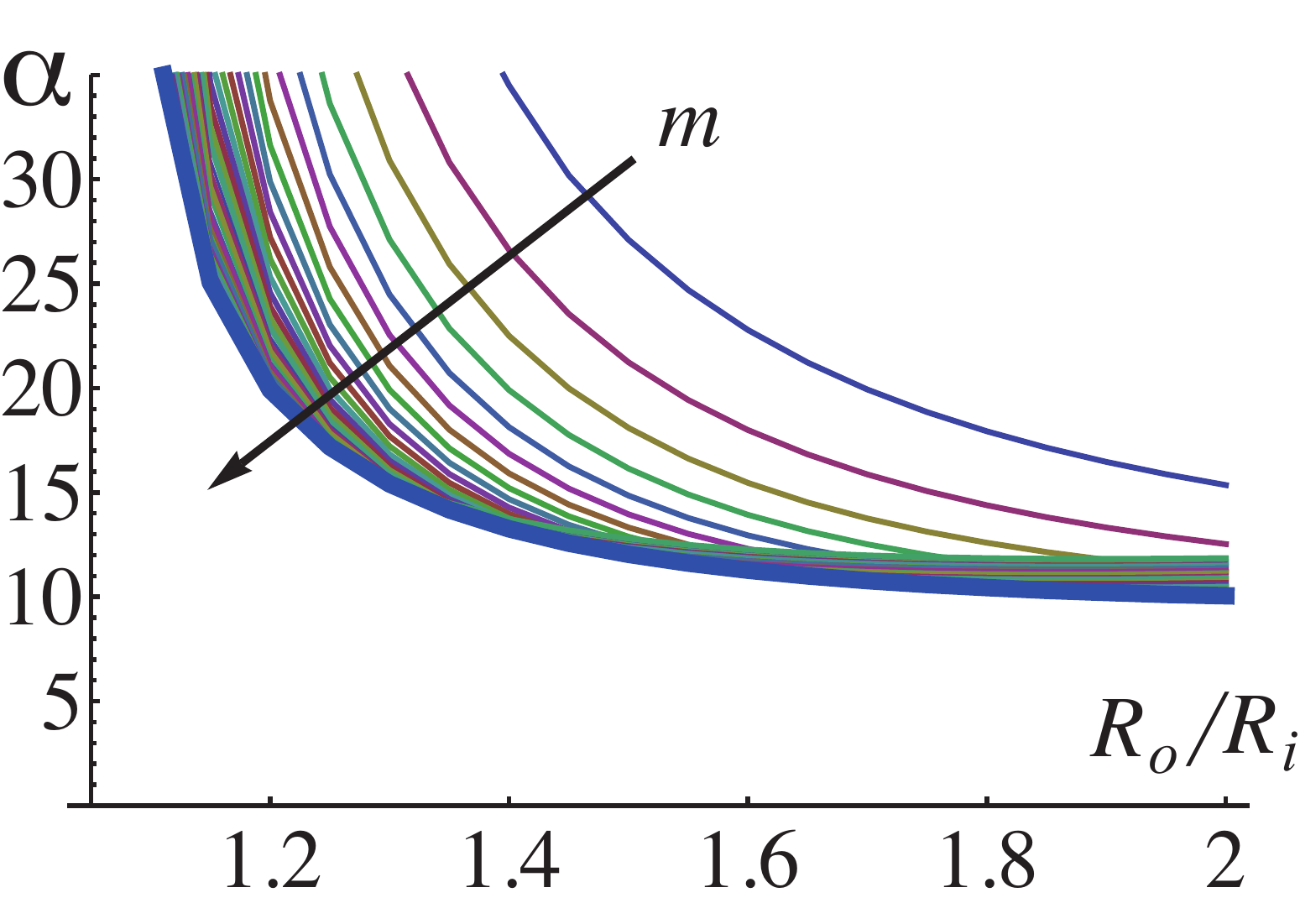}} \quad
\subfigure[compressive residual stress]{\includegraphics[width=0.46\textwidth]{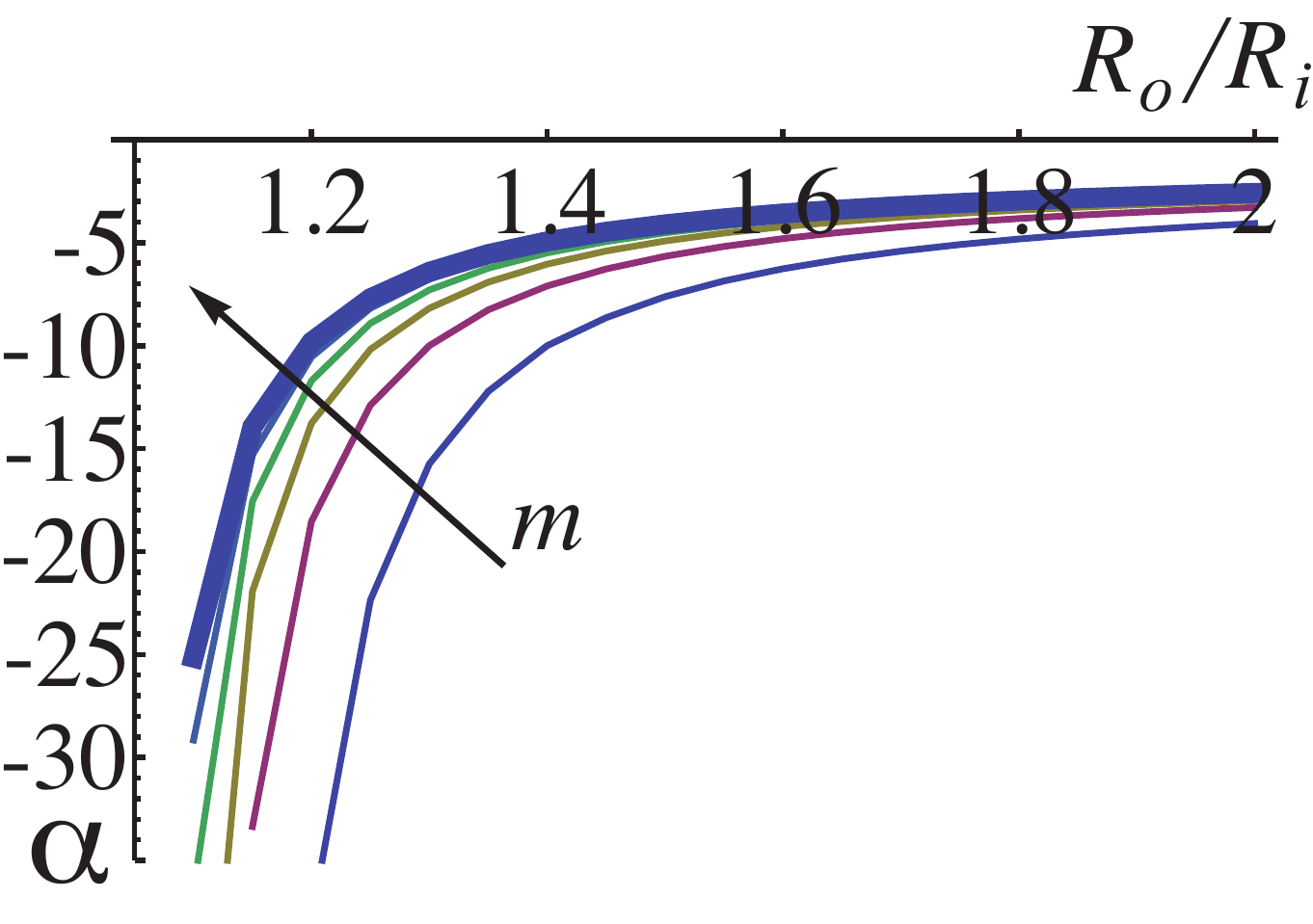}}}
\caption{Instability curves for parabolic radial residual stress ($f(R)$ in Eq.(\ref{fconst}$a$)), when $m=2,3, \ldots, 40$ (left) and when $m=20, 30, \ldots, 70$ (right). 
The thick blue line represents smallest value of $|\alpha|$ for instability to occur versus the aspect ratio $R_o/R_i$ of the tube.}\label{marg-lin}
\end{figure}

\begin{figure}[!htb]
\centering
{\centering
\subfigure{\includegraphics[width=0.55\textwidth]{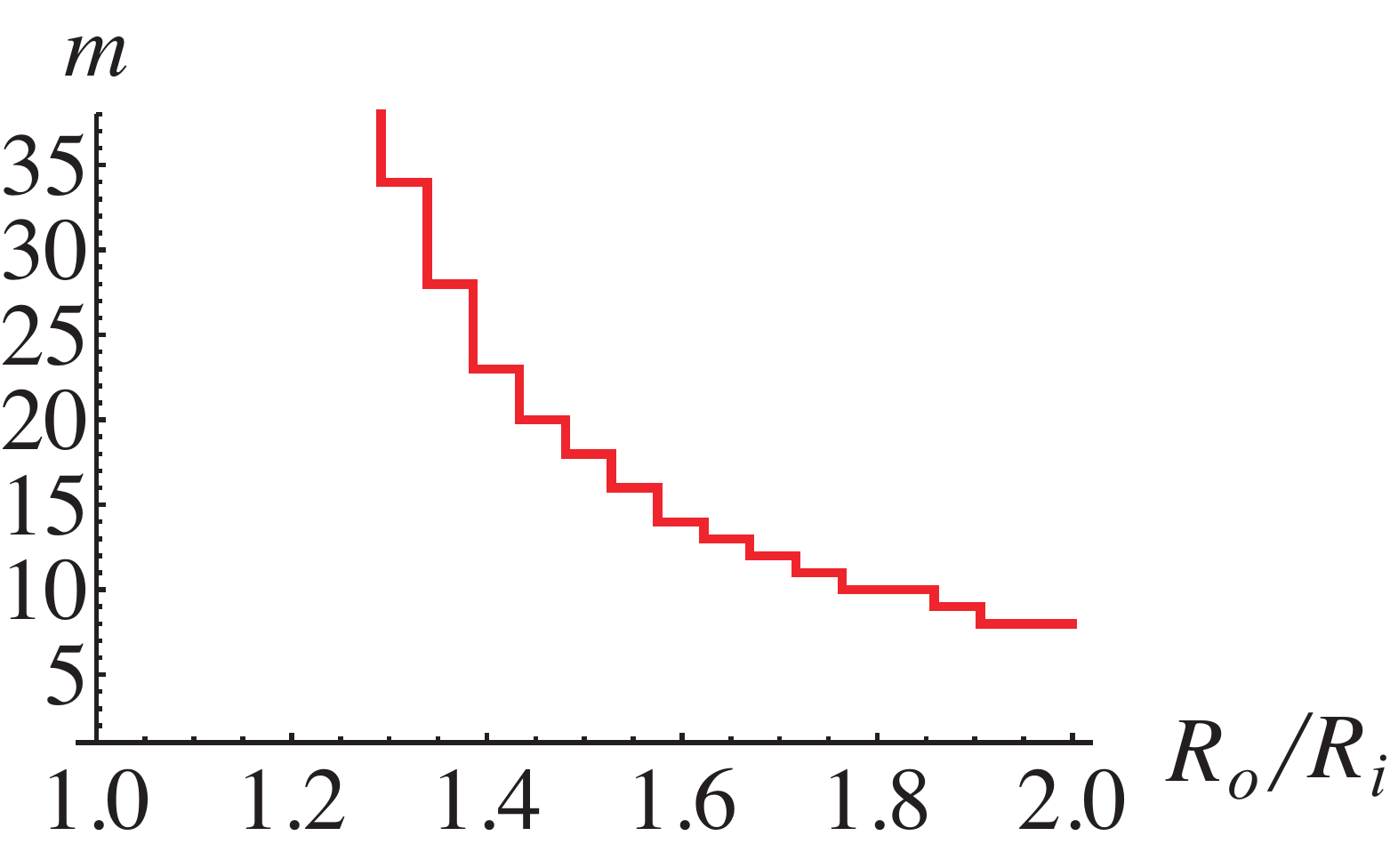}}
\quad
\subfigure{\includegraphics[width=0.4\textwidth]{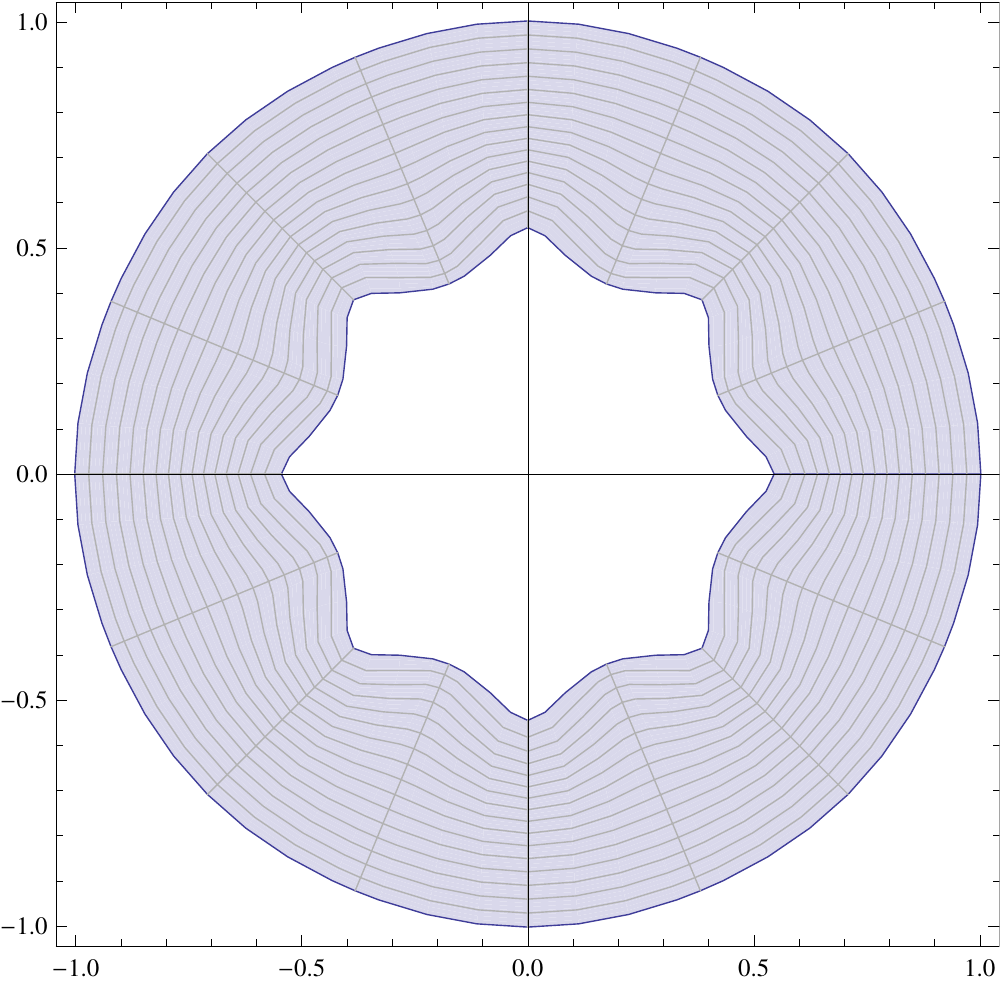}}}
\caption{
Critical circumferential wavenumber $m$  for tensile parabolic residual stress versus the aspect ratio $R_o/R_i$ of the tube (left). 
Solution of the incremental problem for the wrinkled tube (right) at $R_o=1$, $R_i=0.5$; then the critical wavenumber is $m=8$. 
}
\label{k-lin}
\end{figure}

For Case (\ref{fconst}$a$), i.e. $f(R)=\alpha  \mu  R  (R-R_i) (R-R_o)/R_i^2$, the instability curves are depicted for various wavenumbers in Figure \eqref{marg-lin}.
In particular, we find that for both positive and negative values of $\alpha$, the instability curves depend strongly on the aspect ratio of the tube (see Figure \ref{marg-lin}).
The critical wavenumbers for positive $\alpha$ (tensile residual stress) are shown in Figure \ref{k-lin}(a), with the corresponding in-plane wrinkles on the inner face when $R_o/R_i=2$ shown in Figure \ref{k-lin}(b).

\begin{figure}[!htb]
\centering
{\centering
\subfigure[(a)][instability thresholds]{\includegraphics[width=0.48\textwidth]{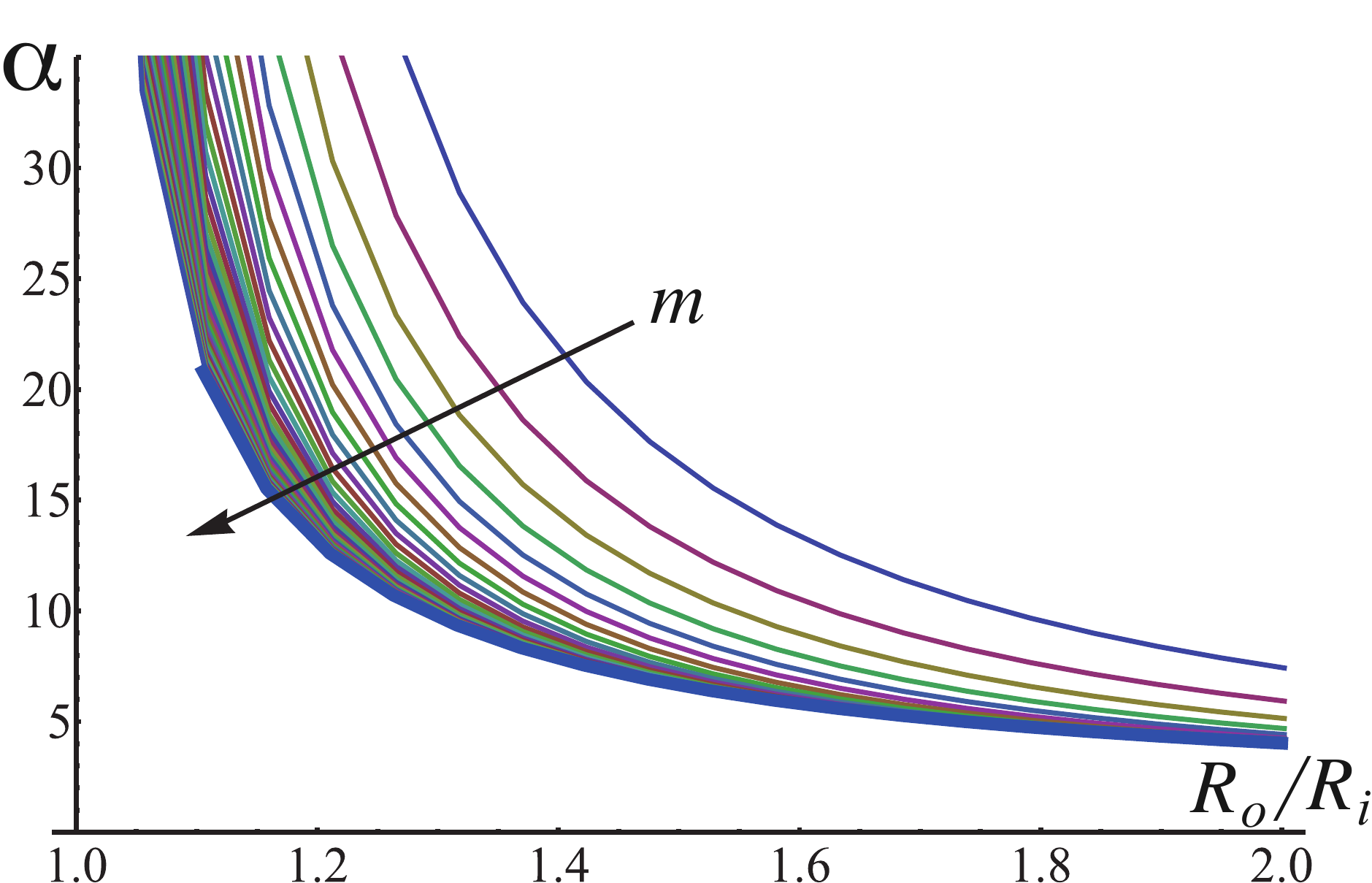}}
\subfigure[(b)][wavenumbers]{\includegraphics[height=5cm]{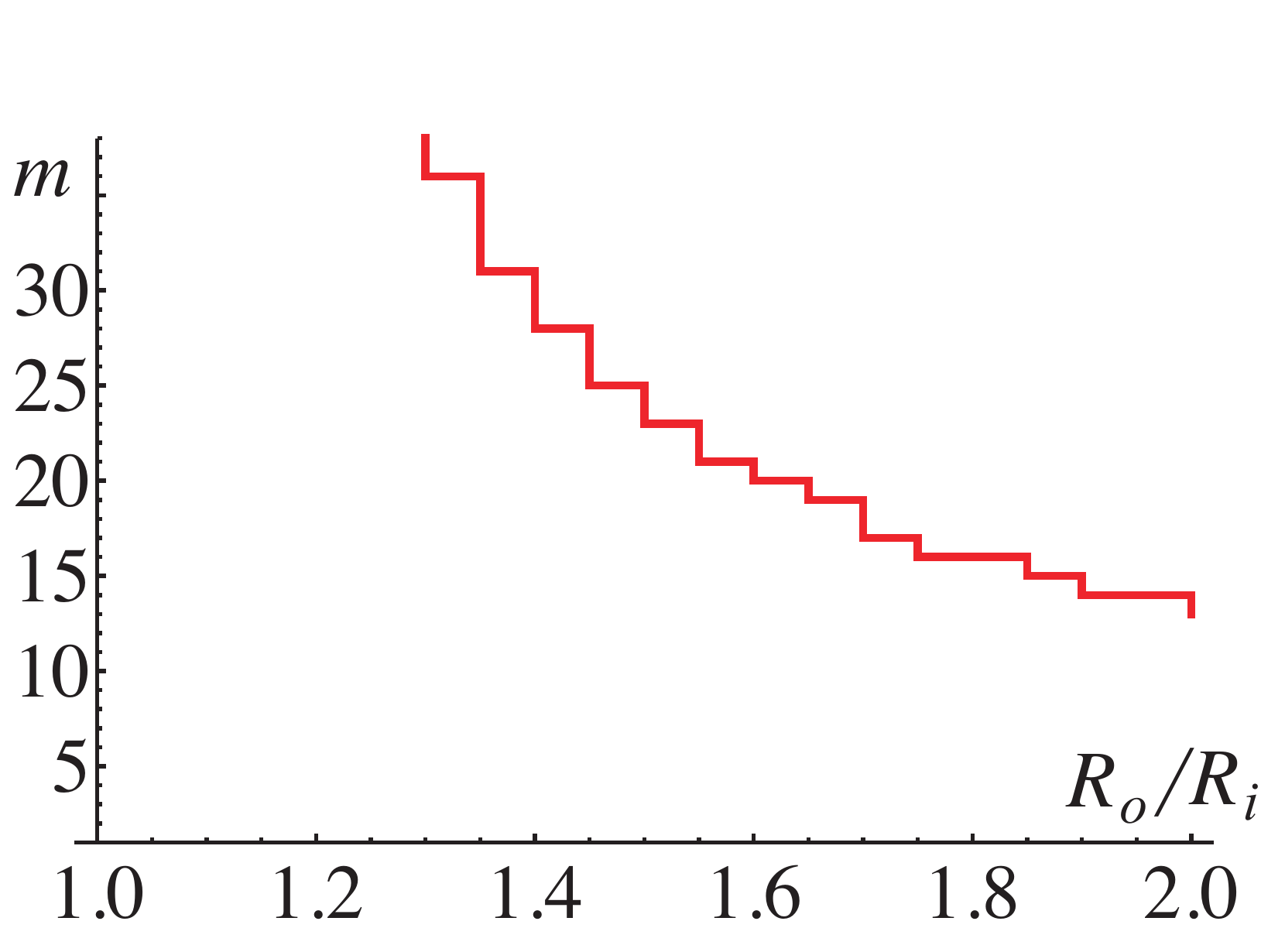}}}
\caption{Instability curves (solid lines) for logarithmic residual stress ($f(R)$ in Eq.(\ref{fconst}$b$)) shown at $m=2,\ldots, 40$. 
The thick blue line represents the critical values of the residual stress amplitude $\alpha$ versus the aspect ratio $R_o/R_i$ of the tube (left). Critical circumferential wavenumber $m$ versus the aspect ratio $R_o/R_i$ of the tube (right).}
\label{marg-log}
\end{figure}

\begin{figure}[!htb]
\centering
{\centering
\subfigure[(a)][tensile residual radial stress]{\includegraphics[width=0.45\textwidth]{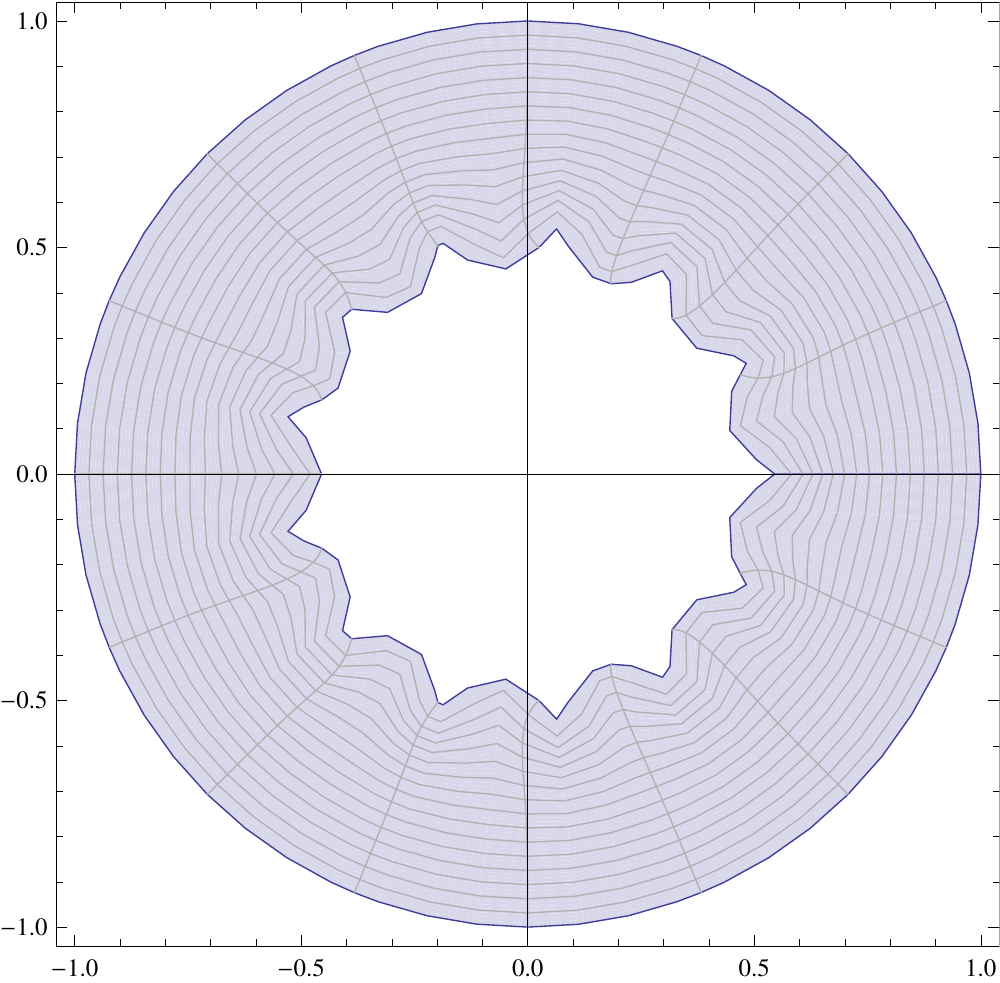}}
\qquad
\subfigure[(b)][compressive residual radial stress]{\includegraphics[width=0.45\textwidth]{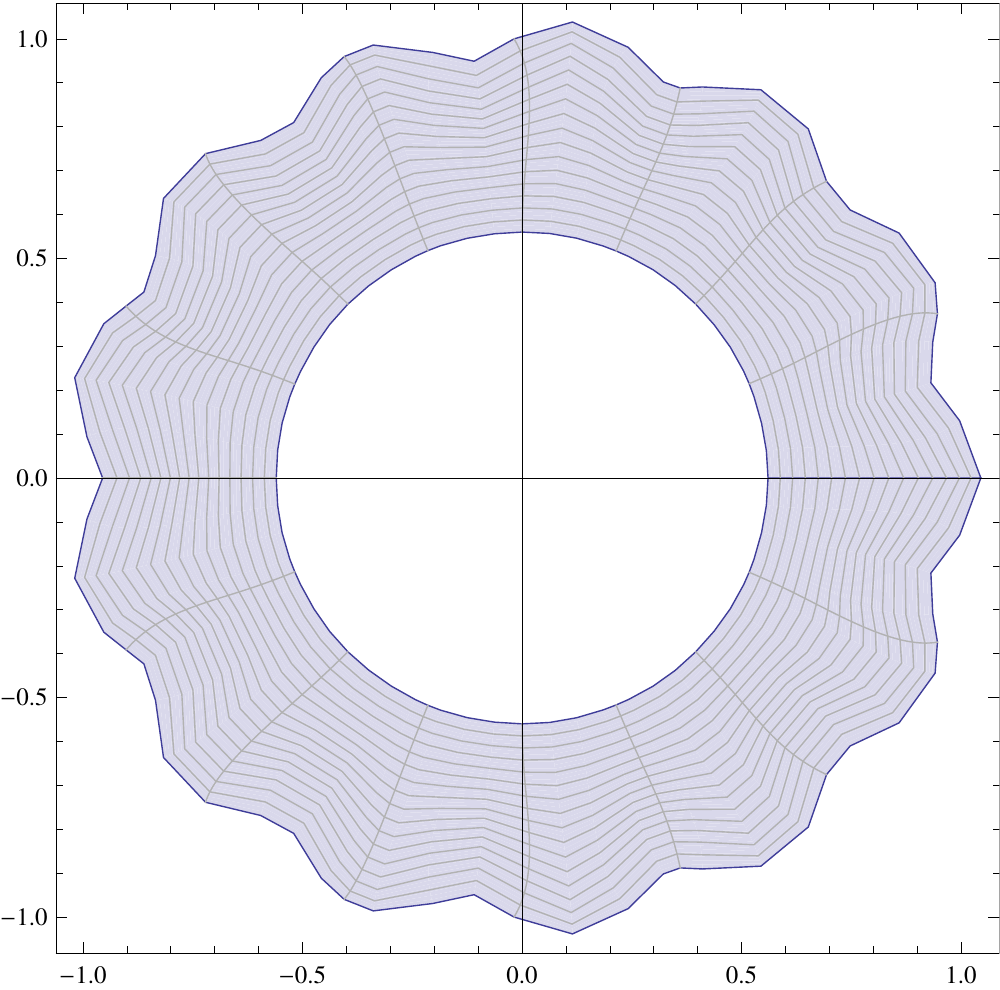}}}
\caption{
Solution of the incremental problem for the tube with logarithmic residual stress ($f(R)$ in Eq.(\ref{fconst}$b$)), when $R_o=1$, $R_i=0.5$, shown for tensile (left) and compressive (right) residual stresses.
Here the critical circumferential wavenumber is $m=13$.}
\label{asp-log}
\end{figure}

For Case (\ref{fconst}$b$), i.e. $ f(R)=\alpha \mu R \ln(R/R_i) \ln(R/R_o)$, the instability curves  occur at the same absolute value of $\alpha$, so that a single figure is required to display the results for both tensile and compressive residual stresses, see Figure \ref{marg-log}(a).
We note that the solution in the tensile case  $\alpha<0$, with its wrinkles on the outer face, is reminiscent of the one proposed by Dervaux and Ben Amar \cite{DeBe11} for the edge buckling of a growing thin ring of gel enclosing a hard disc. 
See Figure \ref{asp-log} for an example of two wrinkled configurations.

\begin{figure}[!htb]
\centering
{\centering
\subfigure[(a)][]{\includegraphics[height=5cm]{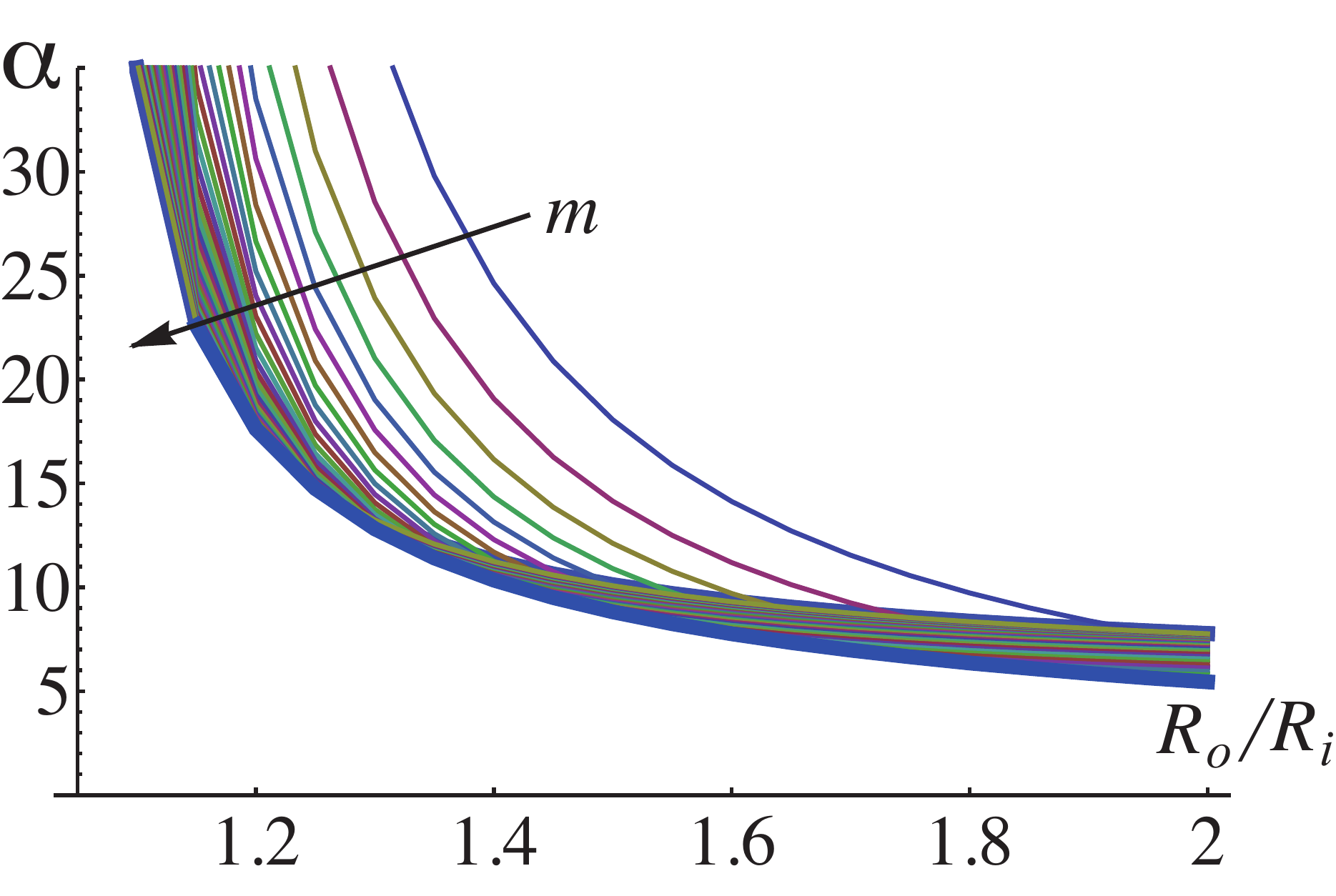}}
\subfigure[(b)][]{\includegraphics[height=4.5cm]{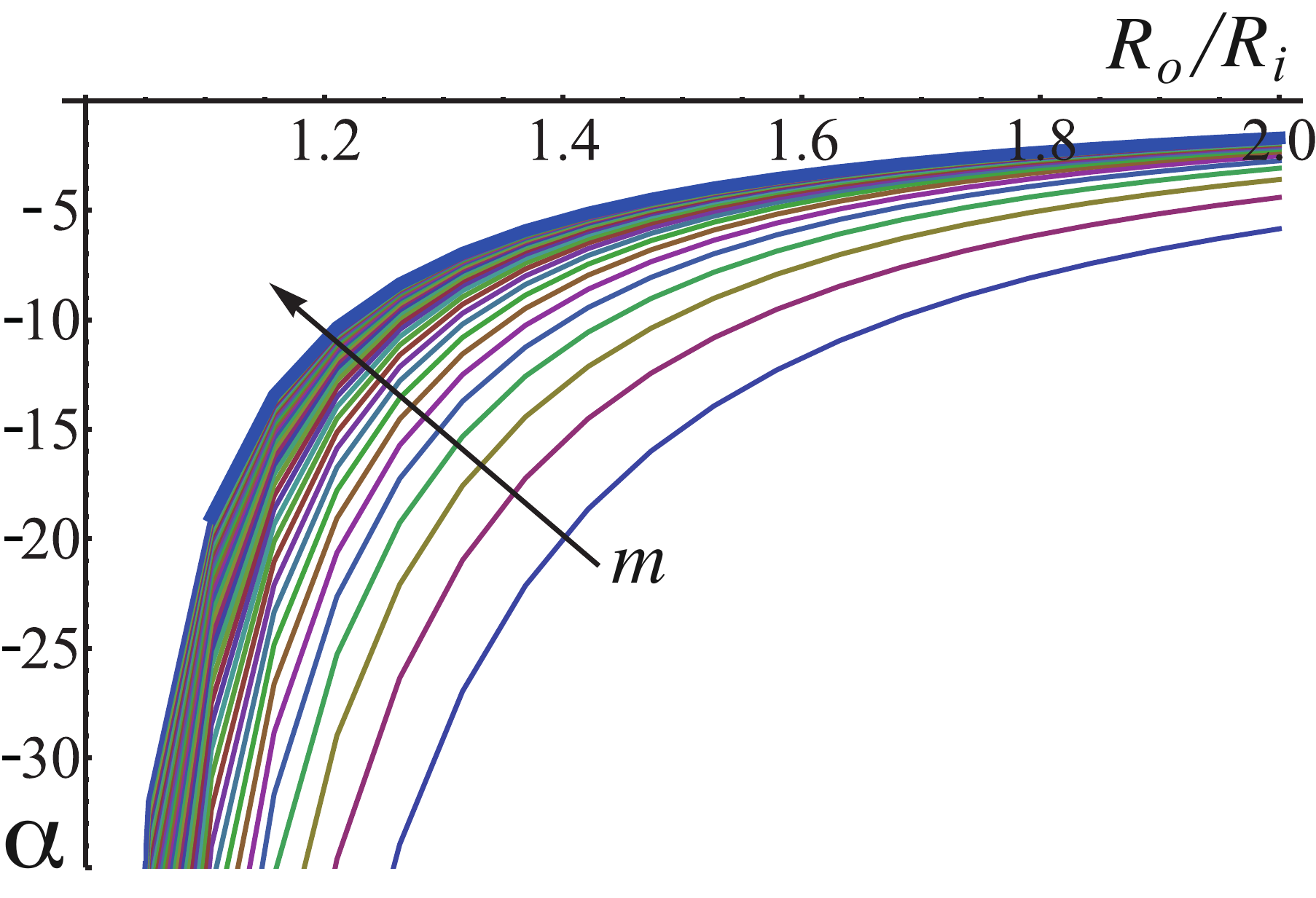}}}
\caption{
Instability curves (solid lines) for exponential residual stress ($f(R)$ in Eq.(\ref{fconst}$c$)), shown at $m=2,\ldots, 40$ for the positive (left) and negative (right) values of the residual stress amplitude $\alpha$. 
The thick blue line represents the lowest critical values of $|\alpha|$ for the onset of wrinkles versus the aspect ratio $R_o/R_i$ of the tube.}\label{marg-exp}.
\end{figure}

\begin{figure}[!htb]
\centering
{\centering
\subfigure{\includegraphics[height=4.9cm]{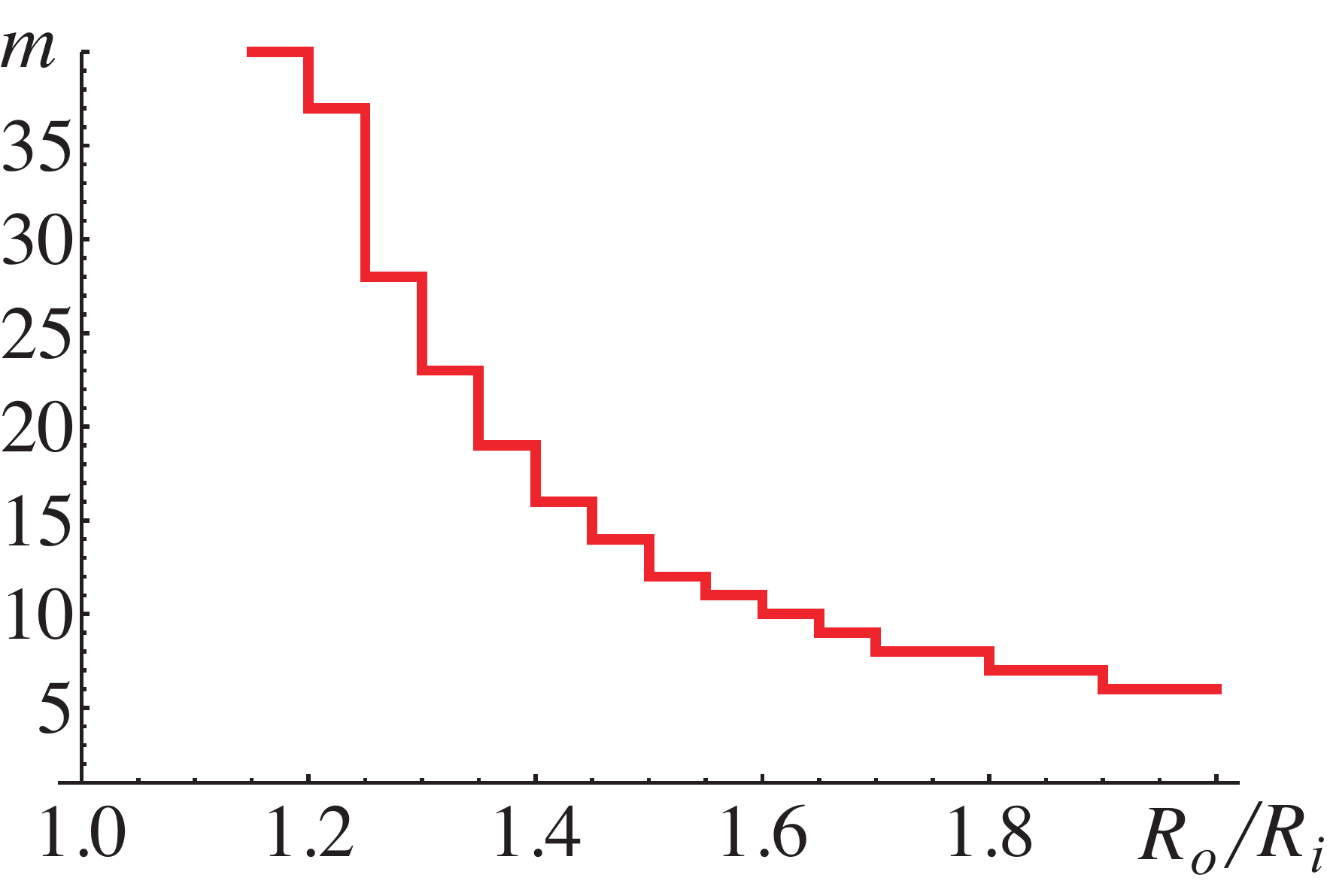}}
\qquad
\subfigure{\includegraphics[height=4.9cm]{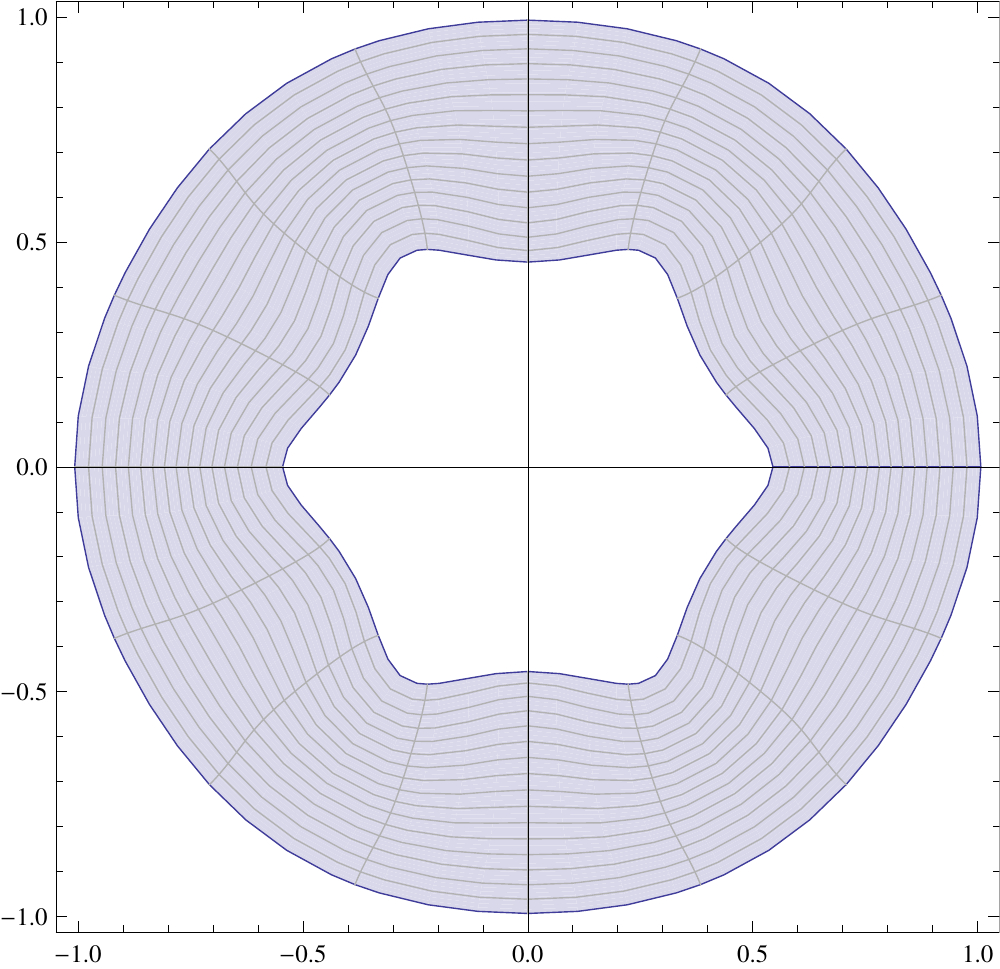}}}
\caption{ 
Critical circumferential wavenumber $m$  for positive values of the critical amplitude $\alpha$, using an exponential stress potential $f(R)$ in Eq.(\ref{fconst}$c$), versus the aspect ratio $R_o/R_i$ of the tube (left). 
Solution of the incremental problem for a tube with $R_o=1$, $R_i=0.5$, and the critical  wavenumber is $m=6$.}
\label{k-exp2}
\end{figure}

Finally, the results for Case (\ref{fconst}$c$), i.e. $ f(R)=\alpha \mu R \left({\rm e}^{R/R_i}-1\right)\left({\rm e}^{R/R_o} - 1\right)$, are qualitatively similar to those of Case (\ref{fconst}a), with instability curves depicted in Figure \ref{marg-exp}.
In particular, the critical circumferential wavenumber $m$ of the instability for the positive values of $\alpha$ is depicted in Figure \ref{k-exp2}  (left), whilst its incremental solution for the wrinkled tube is shown in Figure \ref{k-exp2} (right).

In the next section, we use the critical thresholds calculated by solving the incremental elastic problem as the basis from which to perform a numerical post-buckling investigation of the fully nonlinear morphology of the residually stressed tube.


\section{Numerical post-buckling simulations}
\label{SEC:numeric}


In this Section, we investigate numerically the morphology of  the  residually stressed tube when the parameter $\alpha$, governing the intensity of the residual stress distribution, goes far beyond the linear stability threshold of wrinkling. 
First, we describe the implementation of a numerical finite element method, validating the simulation results with the theoretical predictions for the  linear stability thresholds. 
Second, we investigate the morphology of the residually-stressed tubes in the  post-buckling regime for the three cases of stress potential (parabolic, logarithmic, exponential). 


\subsection{Description of the numerical model}


We implemented the mixed variational formulation of the neo-Hooken model with residual stress Eq.\eqref{engen} into the open source code {FEniCS} \cite{logg2012automated}. 
We considered a hollow cylinder  with $R_{o}=1.0$, $R_{i}=0.5$ as the initial 2D geometry.
We performed quasi-static simulations using triangular {\it Mini} elements: the displacements are discretized with piecewise linear functions enriched by cubic bubble functions, whereas the pressure is discretized by continuous piecewise linear functions.
 
To avoid rigid motions, we imposed null displacements for all the points of the external (internal) boundary in the case of positive (negative) values of $\alpha$. 
We checked {\it a posteriori} the effect of this kinematic constraint compared with the stress-free condition considered theoretically. 
In all numerical simulations we found indeed that  the radial stress is zero at the fixed boundary, even in the post-buckling regime.

For the onset of localized instabilities, we followed Ciarletta \emph{et al.} \cite{ciarletta14} and imposed an initial sinusoidal imperfection with a prescribed mode $m$ and amplitude $0.0025$ on the inner (outer) face nodes for $\alpha$ positive (negative), corresponding to the incremental wrinkles of Section \ref{section3}.
The finite element model was then solved using a incremental iterative Newton-Raphson method over the control parameter $\alpha$, with an automatic adaptation of the step size.


\subsection{Validation versus the theoretical predictions}


We validated the numerical model by comparing it against the stability curves from the theoretical analysis of Section \ref{section3}. 
For this purpose, the numerical thresholds $\alpha_\text{sim}^\text{th}$ were computed as the values of $\alpha$ such that the ratio between the total energy of the system computed numerically, $E_\text{num}$, and theoretically, $E_\text{theo}$, (i.e. for the axi-symmetric solution) is such that $E_\text{num}/E_\text{theo}=0.9999$.
\begin{figure}[!h]
{
\subfigure[(a)][tensile residual stress]{\includegraphics[width=0.45\textwidth]{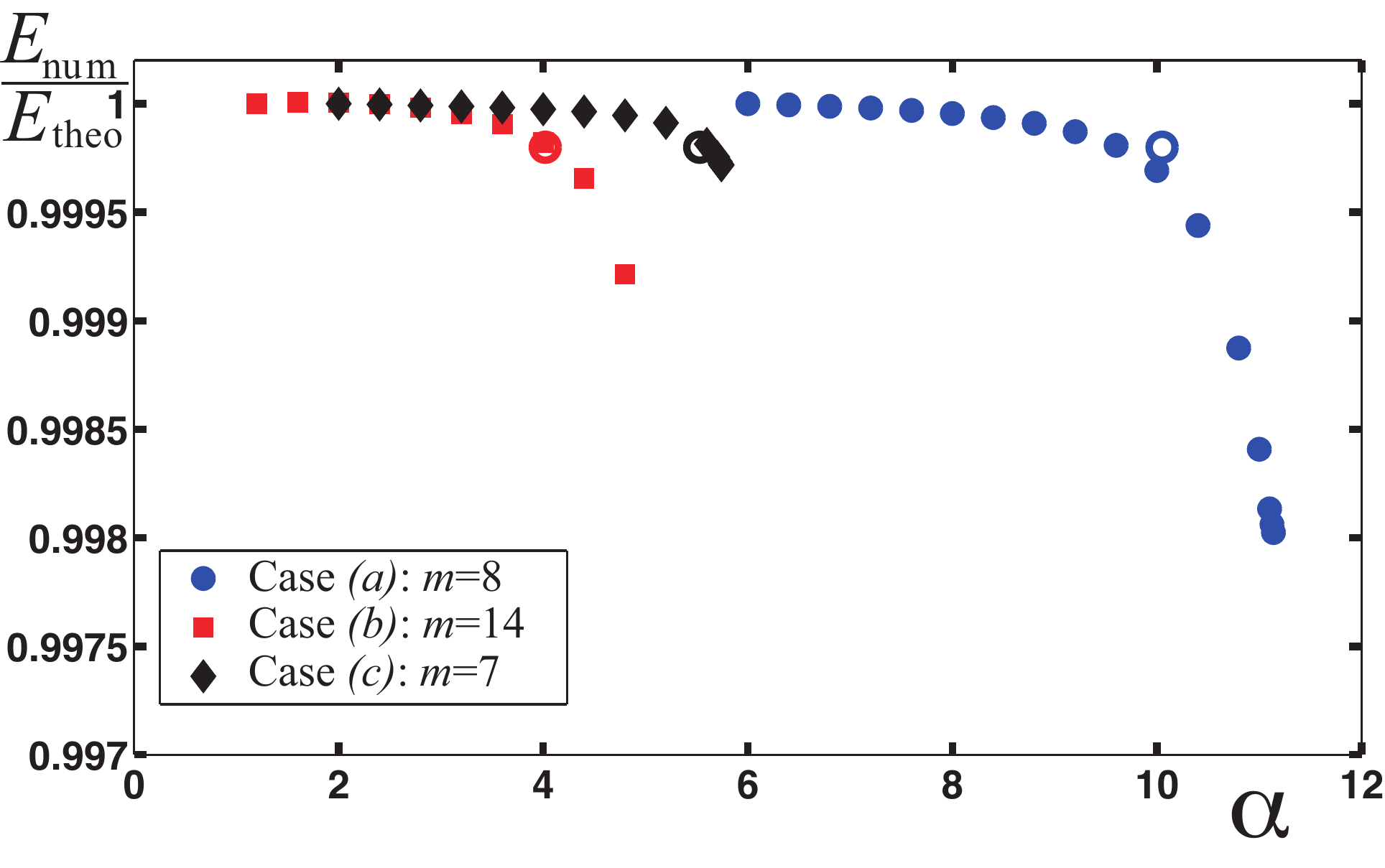}}\quad
\subfigure[(b)][compressive residual stress]{\includegraphics[width=0.45\textwidth]{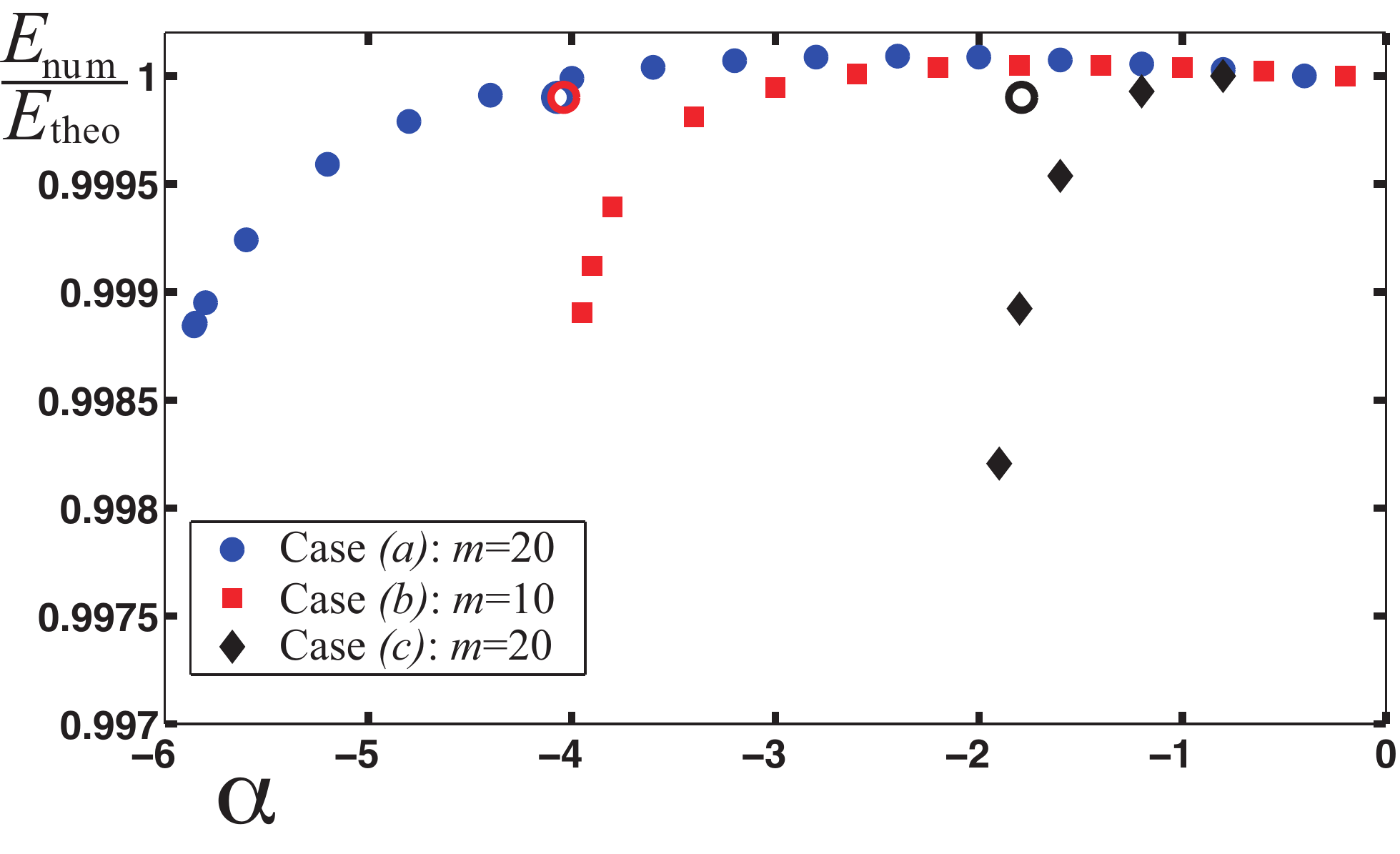}}}
\caption{Energy ratio ${E_{num}}/{E_{theo}}$ versus $\alpha$ for three choices of residual stress fields, when $R_{o}/R_ {i}=2.0$ and $\alpha$ is either positive (a) or negative (b). The hollow circles indicate the theoretical predictions for the stability boundary thresholds ({\bf not sure I understand the threshold bit.}). 
The  theoretical and simulated linear stability thresholds are, in (a): $\alpha_ \text{theo}^ \text{th} = 10.05$; $\alpha_ \text{sim}^ \text{th} = 9.62$ (blue discs); $\alpha_ \text{theo}^ \text{th} = 4.02$; $\alpha_ \text{sim}^ \text{th} = 4.00$ (red squares); $\alpha_ \text{theo}^ \text{th} = 5.53$; $\alpha_ \text{sim}^ \text{th} = 5.55$ (black diamonds); in (b): $\alpha_ \text{theo}^ \text{th} = -4.07$; $\alpha_ \text{sim}^ \text{th} = -4.2$ (blue discs); $\alpha_ \text{theo}^ \text{th} = -4.05$; $\alpha_ \text{sim}^ \text{th} = -3.3$ (red squares); $\alpha_ \text{theo}^ \text{th} = -1.79$; $\alpha_ \text{sim}^ \text{th} = -1.3$ (black diamonds).}
\label{FIG:energy}
\end{figure}

Figure \ref{FIG:energy}(a) depicts  the  ratio $E_\text{num}/E_\text{theo}$ versus $\alpha>0$ for the three cases of residual stress Eq.\eqref{engen}. 
The mode $m$ of the initial imperfection in each simulation has been chosen as the most critical condition predicted theoretically. 
Figure \ref{FIG:energy}(b) shows the energy ratio in function of negative values of $\alpha$. 
Then, the mode $m$ of the initial imperfection has been chosen arbitrarily for the sake of graphical clarity, since the unstable wavelength is very short. 
The theoretical values (indicated by circles in Figure \ref{FIG:energy}) are also reported, highlighting the good agreement of the numerical results with the theoretical predictions. 


\subsection{Residually stressed morphology in the post-buckling regime}


Once it has been validated, we use the proposed numerical method to investigate the morphology of the residually stressed configuration when the order parameter $\alpha$ is far beyond the linear stability threshold.

\begin{figure}[!h]
\centering {\centering
\includegraphics[trim={0 0 0 0},clip,width=0.9\textwidth]{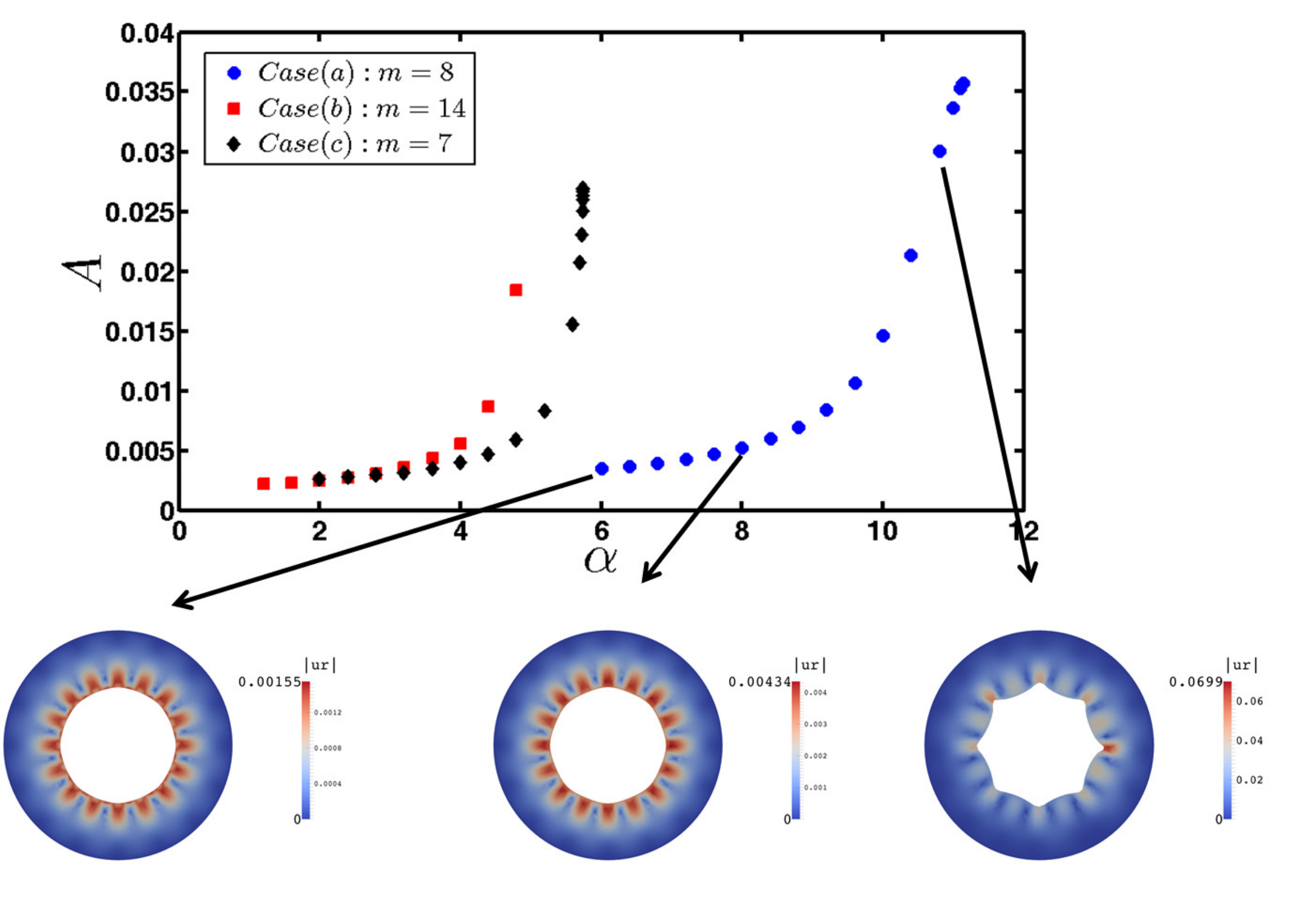}}
\caption {Amplitude evolution of the initial perturbation for the three cases investigated when the control parameter $\alpha$ is positive (tensile residual stress). The deformed morphology is also presented for three indicative values of $\alpha$ in the case of residual stress of parabolic type.}
\label{FIG-morphology}
 \end{figure}

\begin{figure}[!h]
\centering {\centering
\includegraphics[trim={0 0 0 0},clip,width=0.9\textwidth]{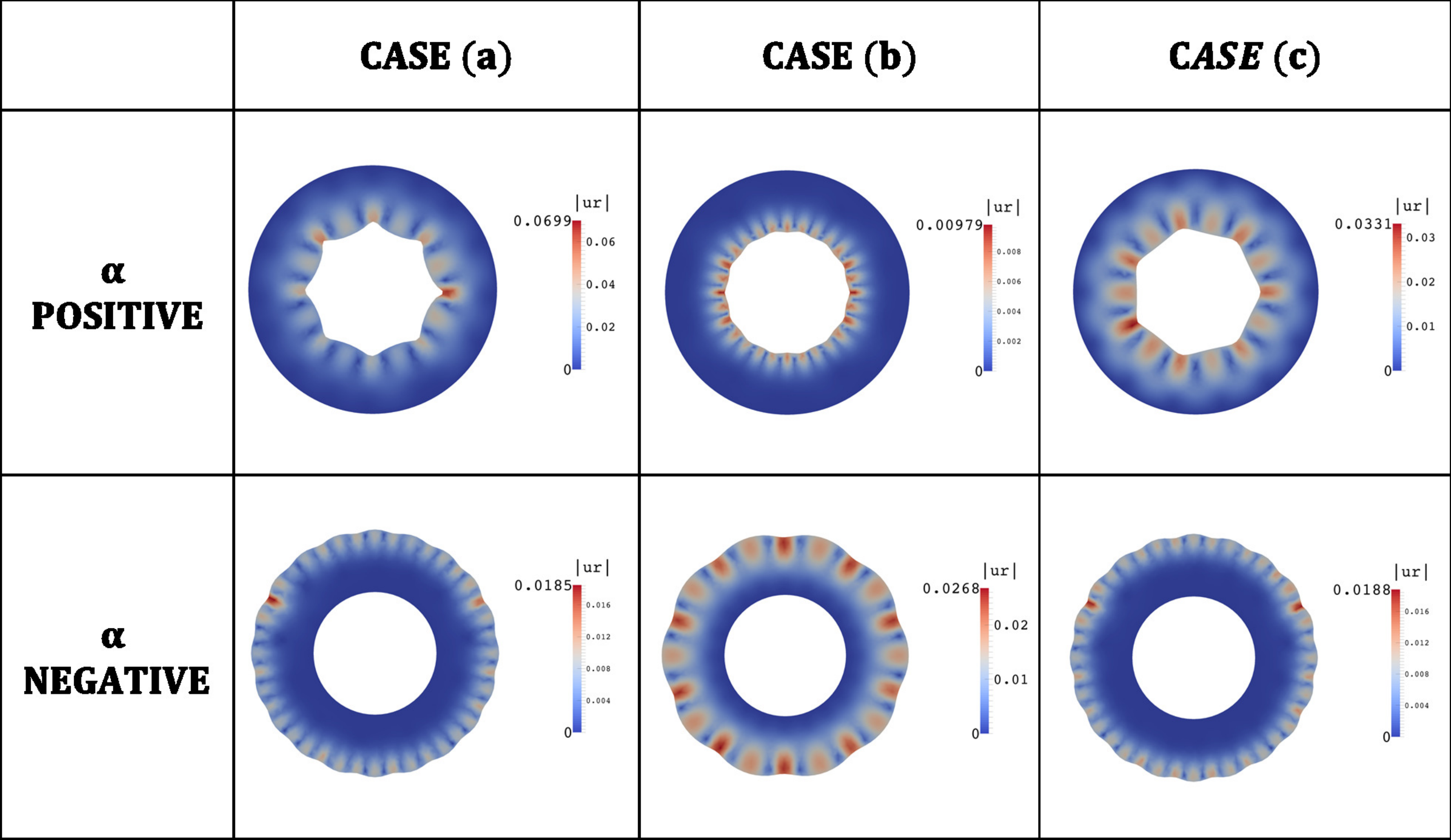}}
\caption {Morphological phase diagram obtained from the finite element simulations as a function of both the underlying distribution of the residual stresses and the positive or negative critical value of $\alpha$. The scalebars indicate the magnitude of the  radial displacement.} 
\label{FIG:deformed}
 \end{figure}

Figure \ref{FIG-morphology} depicts the amplitude $A$ of the  post-buckling  patterns emerging in the numerical simulations for positive values of $\alpha$ in the three cases of residual stress given by Eq.\eqref{fconst}. 
We remark that there is a continuous transition from a wrinkled to a folded pattern, which is indicative of a subcritical bifurcation.
 In Figure  \ref{FIG:deformed} we display the morphological phase diagram for the tube in either positive and negative values of $\alpha$. 

Finally, we remark that the incompressibility constraint for the implemented elements was imposed in a weak sense.  
As a consequence, the morphological transition in the tube can induce a slight shrinkage or enlargement of the boundary elements, eventually creating an excessive distortion and blocking the simulations. 
Thus, future numerical improvements can be concerned with the implementation of an element-wise formulation of the incompressibility constraint.  


\section{Discussion and concluding remarks}
\label{section5}


In this work we used the recently proposed constitutive model of Eq.\eqref{engen} to examine first the equilibrium and next the destabilization of soft tubular materials with increasing levels of residual stresses. 
 
We saw that under plane residual stress, the components of the elastic Cauchy stress tensor can be expressed as a function of a stress potential, which allows a straightforward simplification of the problem in the axi-symmetric case. 
In particular, we focused on three different residual stress distributions: parabolic, logarithmic and exponential radial stress as expressed by Eq.\eqref{fconst}, and we investigated the stability of the resulting configurations using a mix of analytical and numerical techniques.

First, we employed the theory of incremental deformations to study to onset of an elastic bifurcation. 
Using the Stroh formulation and the surface impedance method, the incremental boundary value problem was transformed into the differential Riccati equation \eqref{eqZ} with initial conditions \eqref{bcZ}, which was solved by implementing a robust numerical procedure. 
The resulting  instability curves for the three cases are shown in Figures \eqref{marg-lin}$-$\eqref{k-exp2}, together with the  bifurcated morphologies. 
In particular, we demonstrated that the emerging patterns are localised at the free surface having a threshold compressive value of the hoop residual stress, strongly depending on the ratio between the outer and inner radii. 
These patterns can arise on the inner (in the case of a tensile residual stress) as well as on the outer (in the case of a compressive residual stress) faces of the tube.

Second, we implemented a novel finite element code of the proposed model Eq.\eqref{engen} in order to study the post-buckling evolution of the emerging morphology. 
We validated the numerical code by checking the linear stability thresholds for the onset of buckling, and we derived the post-buckling morphologies. 
In particular, we showed that the accumulation of a compressive hoop residual stress in the fully nonlinear regime drives the transition from a wrinkled to a creased state, which is typically observed in several tubular organs \cite{ciarletta14}. 

The results of this study may have important applications for the non-destructive determination of residual stress distribution in soft tubular tissues. 
Existing approaches aim at deriving first the virtual stress-free state by performing multiple cuttings, possibly an infinity of them (e.g.\cite{lu03, dou06}).
By contrast, our method and results allow to correlate directly the geometrical parameters of the wrinkled or creased morphology with the  spatial distribution of the underlying residual stress components. Thus, we open a novel perspective for guiding the use of non-invasive  techniques to measure residual stresses within living matter.  
Future works will be devoted to improve the constitutive model in order to take into account more complex material behaviors, e.g. describing the  elasticity of real biological networks \cite{storm05}, and structural anisotropy, which is almost ubiquitous in tubular organs \cite{gasser06}.


\section*{Acknowledgements}


Partial funding by the ``Start-up Packages and PhD Program" project, co-funded by Regione Lombardia through the ``Fondo per lo sviluppo e la coesione 2007-2013", formerly FAS program, is gratefully acknowledged. 
So is the support from the Irish Research Council and the EPSRC (reference EP/M026205/1).

\nocite{*}
\bibliographystyle{abbrv}

\end{document}